\documentclass[pdflatex,sn-nature]{sn-jnl}
\usepackage{graphicx}
\usepackage{amsmath}
\usepackage{geometry}
\usepackage{caption}
\usepackage{subcaption}
\usepackage{booktabs}
\usepackage{longtable}
\usepackage[table]{xcolor}
\usepackage{multirow}
\usepackage[percent]{overpic}
\definecolor{humangray}{gray}{0.90}   
\definecolor{vlmblue}{HTML}{1F77B4}   
\definecolor{clipgreen}{HTML}{2CA02C} 
\definecolor{rnlight}{HTML}{FFBB78}   
\newcommand{\rot}[1]{\rotatebox{70}{\strut #1}}

\title{Vision-language models learn the  geometry of human perceptual space}

\author[1]{Craig Sanders}
\author[2]{Billy Dickson}
\author[2]{Sahaj Singh Maini}
\author[1]{Robert Nosofsky}
\author[1,2]{Zoran Tiganj}

\affil[1]{Department of Psychological and Brain Sciences, Indiana University Bloomington}
\affil[2]{Department of Computer Science, Indiana University Bloomington}

\begin{document}
\maketitle

\begin{center}\large\bfseries Abstract\end{center}
\vspace{-0.5\baselineskip}

\begin{abstract}\leavevmode
\noindent In cognitive science and AI, a longstanding question is whether machines learn representations that align with those of the human mind. While current models show promise, it remains an open question whether this alignment is superficial or reflects a deeper correspondence in the underlying dimensions of representation. Here we introduce a methodology to probe the internal geometry of vision-language models (VLMs) by having them generate pairwise similarity judgments for a complex set of natural objects. Using multidimensional scaling, we recover low-dimensional psychological spaces and find that their axes show a strong correspondence with the principal axes of human perceptual space. Critically, when this AI-derived representational geometry is used as the input to a classic exemplar model of categorization, it predicts human classification behavior more accurately than a space constructed from human judgments themselves. This suggests that VLMs can capture an idealized or `denoised' form of human perceptual structure. Our work provides a scalable method to overcome a measurement bottleneck in cognitive science and demonstrates that foundation models can learn a representational geometry that is functionally relevant for modeling key aspects of human cognition, such as categorization. 
\end{abstract}

\keywords{Vision-language models, Psychological space, Categorization, Cognitive modeling, Human-AI alignment}

\section{Introduction}

A central aim in cognitive science is to explain how humans carve high-dimensional visual input into psychologically meaningful structure that supports similarity, generalization, and categorization \cite{Rosch1975,Medin1978}. A powerful approach is to model behavior in terms of distances in a psychological space, where nearby items are judged as similar and generalization falls off with distance \cite{Shepard1987,Tversky1977,RoadsLove2024ARP}. Multidimensional scaling (MDS) provides a classic route to such spaces by embedding (dis)similarity data into low-dimensional Euclidean configurations \cite{Kruskal1964a,Groenen2016}. These embeddings have underpinned quantitative theories of categorization, most prominently exemplar-based accounts, such as the Generalized Context Model (GCM), which predict category choice probabilities from similarities of stimuli to stored exemplars \cite{Nosofsky1986}. However, progress has been hampered by a fundamental measurement problem: collecting large-scale human similarity matrices remains a practical bottleneck \cite{Goldstone1994,Hout2013,Richie2020,Hebart2023} --  exhaustive pairwise ratings grow quadratically with the number of items and are burdensome to acquire \cite{Goldstone1994,Hout2013,Richie2020}. Spatial-arrangement methods, multiple-query procedures, and large curated resources offer partial relief \cite{Hout2013, Hebart2023,RoadsLove2021HSJ,roads2019obtaining}, but comprehensive similarity matrices for naturalistic domains remain scarce. Notably, recent work has established well-validated natural-science image sets (e.g., rock types as formalized in the geologic sciences) and corresponding psychological spaces that support formal modeling of category learning \cite{Nosofsky2018BRM,Nosofsky2018JEPG,Nosofsky2019PBR,Nosofsky2020CBB}.

In parallel, advances in vision and multimodal machine learning have yielded representations that correlate with human perceptual judgments and neural responses \cite{Battleday2021,Radford2021CLIP,Yamins2014,KhalighRazavi2014,Cichy2016,Haxby2014,hasan2025training,doerig2025high}. This raises a central question: do large-scale models and humans rely on the same underlying perceptual dimensions? While classic results have shown important divergences, such as a texture bias in ImageNet-trained CNNs compared with humans' stronger reliance on shape \cite{geirhos2018imagenet,Geirhos2020}, recent work suggests that multimodality and language supervision can modulate such biases \cite{VLMShapeBias2024,wang2023better}. Much of the research comparing AI and human representations has focused on establishing broad structural correspondence \cite{RSA2008,Kriegeskorte2013Geometry,Nili2014Toolbox,Haxby2014}, for example by correlating entire similarity matrices \cite{Devereux2013RSA,Storrs2021JOCN,Hebart2020NHB,ogg2025turingRSA} or using high-dimensional model embeddings as feature spaces for predicting behavioral data \cite{Peterson2018Correspondence,Battleday2017Categorization,Attarian2020Transforming,Tarigopula2023NeuralNetworks,demircan2024evaluating}. However, limiting analysis to such global correlations can underestimate the true degree of correspondence \cite{Diedrichsen2017,Walther2016}. Humans and models may rely on the same underlying perceptual dimensions yet assign them different relative weights when forming similarity judgments. As a result, their overall matrix correlation may appear low even when coordinate values along individual dimensions are highly aligned. This possibility motivates the need to test for alignment at the level of specific perceptual axes \cite{Hebart2020NHB,Love2021Similarity,RoadsLove2024Dimensions,mahner2025dimensions}. Recent work highlights the potential of using vision-language models (VLMs) as scalable surrogates for human judgments  \cite{Lu2022,dickson2025comparing,marjieh2024large}. A critical gap remains in understanding whether the specific, interpretable axes that structure human psychological space, such as lightness, texture, or color, can themselves be recovered from modern foundation models.

\begin{figure*}[ht!]
			\begin{tabular}{l}
				\textbf{A} \\
				\includegraphics[width=\linewidth]{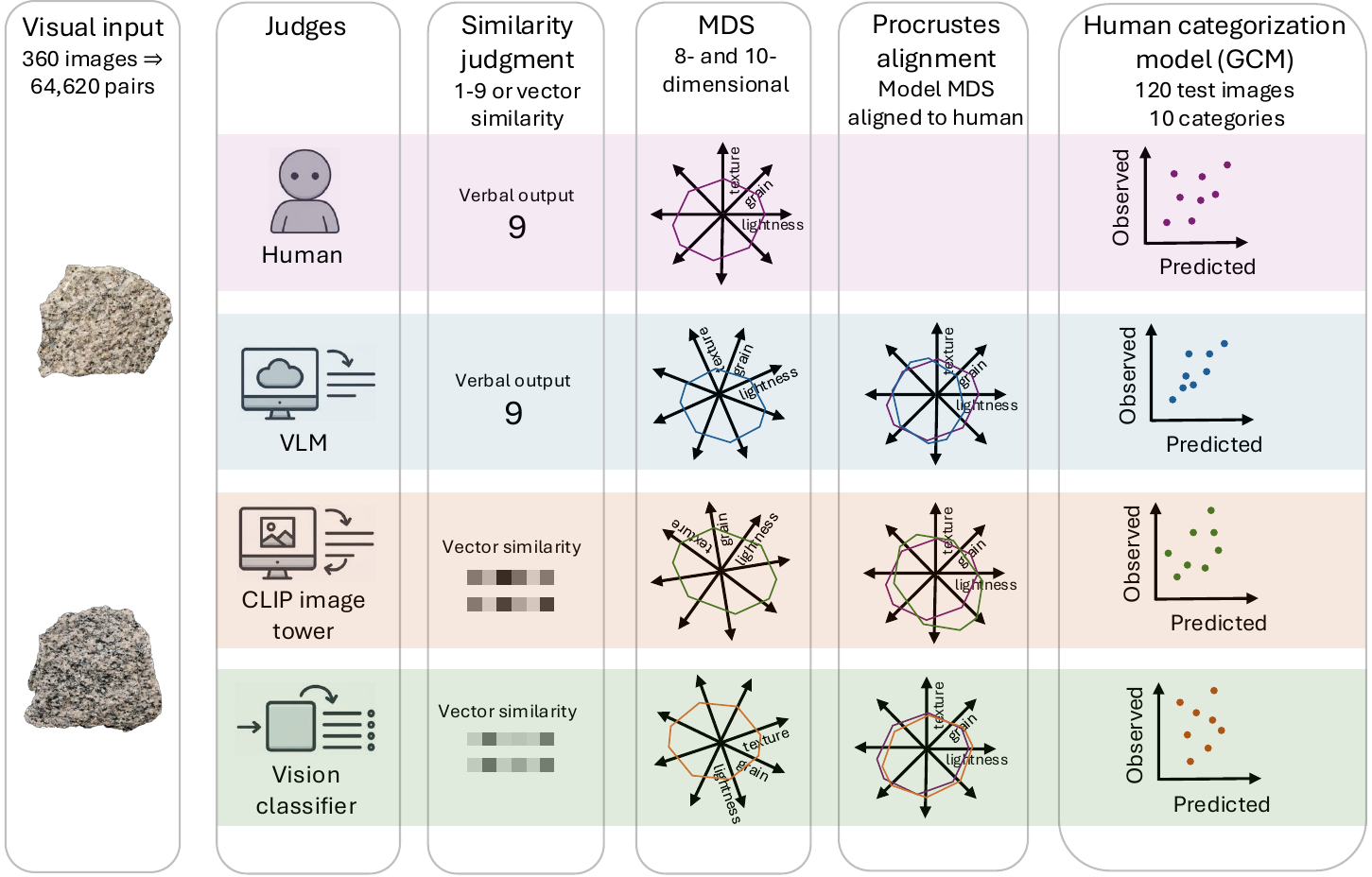} 
			\end{tabular}
            \begin{tabular}{l}
				\textbf{B} \\
				\ \ \ \ \ \ \ \ \ \includegraphics[width=0.85\textwidth]{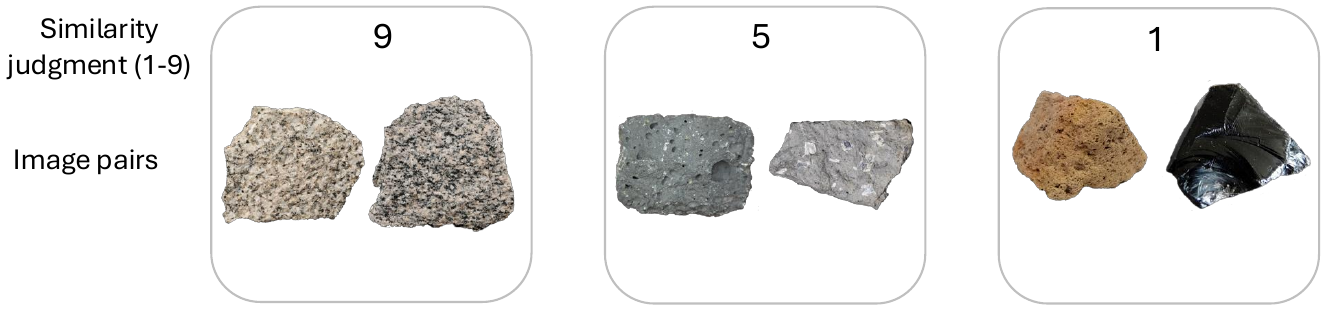}
			\end{tabular}
  \caption{From pairwise similarity to human-like psychological spaces and human categorization prediction.
  \textbf{A.} \textbf{Pipeline.} We use a dataset of 360 rock images and for each pair (64{,}620 pairs), four types of judges provide a similarity measure: humans and VLMs output 1-9 ratings, while frozen CLIP image towers and vision classifiers (ResNet/ViT) yield embedding vectors from which we compute pairwise Euclidean distances. The resulting 360 \(\times\) 360 matrices are embedded with MDS to obtain low-dimensional spaces. Structural validity is assessed via Procrustes alignment of each model space to human normative perceptual axes and to the human 8D MDS. Behavioral validity is assessed by supplying the coordinates to an exemplar GCM to predict human categorization for an independent dataset of 120 test images from 10 categories.
  \textbf{B.} \textbf{Similarity examples.} Three image pairs illustrate the 9-point similarity scale used by humans and the VLM (9 very similar, 5 moderately similar, 1 very dissimilar).}
  \label{fig:schematic}
\end{figure*}

Here, we bridge this gap by introducing a methodology to explicitly test the dimensional alignment between VLM and human perception. We go beyond showing that similarity ratings are correlated and ask whether we can recover a low-dimensional space from a VLM whose axes correspond directly to human perceptual dimensions. We elicit pairwise visual-similarity judgments for rock images from \emph{VLMs}, and compare them to similarity proxies derived from \emph{contrastive image encoders} (CLIP-style towers with ViT/ResNet backbones) and \emph{supervised vision classifiers} (ViT/ResNet). Judgments or distances are transformed into psychological spaces using nonmetric MDS \cite{Kruskal1964a,BorgGroenen2005}. We then assess two criteria. First, \emph{structural validity}: using Procrustes alignment \cite{Gower1975}, we test for a one-to-one correspondence between the dimensions of the model-derived spaces and normative human perceptual dimensions. Second, \emph{behavioral validity}: we test whether these VLM-derived spaces, with their newly identified dimensions, are sufficient to predict human classification probabilities when used as input to an exemplar model (GCM) \cite{Nosofsky2018JEPG,Nosofsky2020CBB}. An overall schematic of our approach is provided in Figure~\ref{fig:schematic}.

Across model families, large VLMs yield a similarity structure whose underlying dimensions are most closely aligned with human perceptual axes. Procrustes analyses reveal a striking correspondence for canonical perceptual dimensions like lightness, grain size, and shininess. Behaviorally, when these AI-derived psychological spaces are used as the representational substrate for an exemplar model, they exceed the predictive utility of spaces built from human similarity ratings themselves. Taken together, these findings demonstrate that modern VLMs not only approximate human similarity structure but do so by converging on a human-like perceptual geometry. This provides both a practical route to measurement for cognitive science and strong evidence for shared constraints on perceptual organization across very different learning systems.

\section{Methods}

\subsection{Stimuli}
The stimuli were digital images of rocks from a dataset previously used in studies of human categorization \cite{Nosofsky2018BRM}. The dataset consists of 360 rock images, comprising 12 samples from each of 30 distinct geological categories. 

\subsection{Human data}
\label{sec:human_data}
We use two sources of human behavioral data as benchmarks and targets for modeling.  
First, prior work collected pairwise similarity judgments for the same rock-image corpus and reported an 8D  MDS solution, along with normative ratings on seven perceptual attributes (lightness/darkness, grain size, rough/smooth texture, shininess, organization, chromaticity, red/green hue) \cite{Nosofsky2018BRM}. In the present study, these human-derived representations serve as (i) a structural reference for comparing model-derived similarity structure and (ii) anchors for Procrustes alignment and dimension-wise interpretability analyses. 
Second, we evaluate behavioral relevance using classification data where human learners classified 120 rock images (a subset of our image corpus) into 10 categories; we use the observed item\(\times\)category response probabilities from that experiment as the target for GCM fits \cite{Nosofsky2020CBB}.

\subsection{Vision-language rater models}
We used a set of VLMs to generate visual similarity ratings for pairs of rock images. For each pair, we provided the model with the two images and a prompt requesting a similarity rating on a 9-point scale. We tested two different prompts. The `baseline' prompt was:
\small{\begin{verbatim}
<System>: You are assisting in a study in which you are shown pairs of rocks 
and rate how visually similar they are on a scale from 1 to 9, with 1 being 
most dissimilar, 5 being moderately similar, and 9 being most similar. 
You only respond with a single number from 1 to 9, without explaining 
your reasoning.
<User>: From 1-9, how visually similar are these two rocks? 
{JPG of Rock1} {JPG of Rock2}
\end{verbatim}}
For GPT-4o model, we observed that this prompt yielded a distribution of ratings skewed toward the low end of the scale compared to human ratings (Figure \ref{fig:rating_histograms}), thus we designed an `encourage middle' prompt to encourage use of the full range of the scale (this prompt was used only with GPT-4o model as part of an exploratory analysis to examine whether prompt engineering can shape the distribution of the model's ratings):
\small{\begin{verbatim}
<System>: You are assisting in a study in which you are shown pairs of rocks and rate 
how visually similar they are on a scale from 1 to 9, with 1 being most 
dissimilar, 5 being moderately similar, and 9 being most similar. You only 
respond with a single number from 1 to 9, without explaining your reasoning. 
You use the full range of the 1-9 scale and err on the side of using the 
middle of the scale (4 or 5) when you are unsure. Rocks that are very 
similar in almost all respects should be rated as highly similar (8 or 9). 
You only use a 1 or 2 when the rocks are truly different in every meaningful 
visual way. Most ratings should fall somewhere in the middle of the scale.
\end{verbatim}}

For each prompt, we obtained similarity ratings for all 64,620 image pairs. For GPT-4o, we retrieved the top 20 candidate responses and their associated log-probabilities from the API, and computed each pair’s final rating as a log-probability-weighted mean. For all other models, we requested a single integer rating directly, as log-probabilities were not exposed through the Ollama API.

We evaluated both locally executed and API-accessed VLMs. For local evaluation, we used Ollama \cite{ollama2025} to run models on 4 NVIDIA H100 GPUs: \texttt{llama4:17b-scout-16e-instruct-fp16} (Mixture-of-Experts; 109B total; 17B active; 16 experts) \cite{Meta2024Llama4Multimodal}, \texttt{qwen2.5vl:\{72b/32b/7b/3b\}-fp16} \cite{Qwen2.5-VL_technical_report}, and \texttt{gemma3:\{27b/12b/4b\}-it-fp16} \cite{gemma3_technical_report}. GPT-4o \cite{OpenAI2023GPT4, OpenAI2024GPT4o} was accessed via API (parameter count not publicly disclosed).

\subsection{Fixed-embedding baselines (CLIP and vision classifiers)}
In addition to VLMs, we used fixed-feature baselines that do not produce token probabilities and where individual stimulus images were used as inputs:
(i) CLIP image towers used in frozen mode \cite{Radford2021CLIP}, and
(ii) vision-only classifiers trained for recognition (ResNet-50/101 \cite{He2016ResNet}, ViT variants \cite{Dosovitskiy2021ViT}).

For CLIP, each stimulus image was passed through the image tower (no text/caption input) and we used the model’s standard image embedding (post-projection).
For vision classifiers, we extracted feature vectors from the network trunk immediately before the classification head (e.g., global-average-pooled features for ResNets; the [CLS] token or pre-head representation for ViTs).

Pairwise Euclidean distances between embeddings yielded a $360 \times 360$ distance matrix per model.
To express geometry on a similarity scale comparable to human matrices, we applied the fixed monotone mapping
$s(d) = \exp(-c d).$
Because nonmetric MDS depends only on the rank order of pairwise
(dis)similarities, our results are robust to the choice of~$c$: after selecting an appropriate order of magnitude 
($c \approx 0.1$ or $1$) to avoid saturation,
further adjustments of~$c$ within that range had negligible effect.

\subsection{Similarity-Rating Analyses and Multidimensional Scaling (MDS)}
For category-level analyses, we formed a 30×30 \textit{category-level} similarity matrix per source by averaging pairwise similarities over all item pairs between each category pair (including within-category entries), then vectorized the upper triangle and computed Pearson’s r between sources. 

We performed nonmetric MDS on the \textit{individual-item-level} similarity matrices to derive spatial representations of the \textit{individual} rock stimuli. The analyses were conducted for solutions of varying dimensionality (from 2 to 12 dimensions). We used Kruskal’s Stress-1 as the measure of goodness-of-fit. The primary analyses focused on 8D and 10D solutions to allow for direct comparison with prior work on human-derived representations \cite{Nosofsky2018BRM,Nosofsky2020CBB}.

\subsection{Procrustes Analysis}
To compare the MDS solutions derived from models with those derived from human data, we used Procrustes analysis. This method finds an optimal affine transformation (rotation, reflection, translation, and dimension-wise scaling) to align one configuration of points onto another, minimizing the sum of squared errors between corresponding points. We rotated the MDS solutions from each of the models to align with: (1) a set of seven normative perceptual dimension ratings collected from human participants (lightness/darkness, grain size, smooth/rough texture, shininess, organization, chromaticity, and red/green hue), and (2) the 8D MDS solution derived from human similarity ratings \cite{Nosofsky2018BRM}. Although normative ratings were not collected for the eighth dimension, the authors reported that it appeared to have shape-related components.

\subsection{Modeling Human Classification Performance}
\label{sec:GCM}
To test the utility of the model-derived psychological space, we used it to predict human performance in a classification task. The data were from a `coverage' condition reported by  Nosofsky et al.\ \cite{Nosofsky2020CBB}, where participants learned to classify 120 igneous rocks into 10 categories. We used the Generalized Context Model (GCM) \cite{Nosofsky1986}, an exemplar-based global-matching model (see \cite{osthglobal} for review), to predict the full 120 (test items) $\times$ 10 (categories) confusion matrix from the experiment.

Under GCM, the probability that a test item $i$ is assigned to category $J$ is proportional to the total similarity of $i$ to the training exemplars of $J$, normalized by the total similarity of $i$ to all exemplars across all $M$ categories (here $M=10$). We also include a global lapse/guess rate $\varepsilon$ that mixes in uniform responding. Specifically, with response-scaling parameter $\gamma$,
\begin{equation}
P(J \mid i)
= (1-\varepsilon)\,
\frac{\big(\sum_{j \in J} s_{ij}\big)^{\gamma}}
     {\sum_{K=1}^{M} \big(\sum_{k \in K} s_{ik}\big)^{\gamma}}
+ \frac{\varepsilon}{M},
\end{equation}
where $s_{ij}=\exp(-c\,d_{ij})$ with sensitivity $c$, and distances are Euclidean in the embedding, \[d_{ij}=\sqrt{\sum_{m=1}^{8} (x_{im}-x_{jm})^{2}}\].

\noindent This `core' GCM therefore has only three free parameters fitting 1200 data points: the lapse/guess rate $\varepsilon$,
the response-scaling parameter $\gamma$, and the sensitivity $c$.

We also tested an extended GCM that incorporated five additional `supplementary' dimensions found to be diagnostic for classification \cite{Nosofsky2020CBB}. In this version, the distance function is:
\[
d_{ij} = \sqrt{\sum_{m=1}^{8} w_{m}(x_{im} - x_{jm})^2 + \sum_{m'=1}^{5} w_{m'} (x'_{im'} - x'_{jm'})^2},\]
where $x'_{im'}$ are the ratings on the supplementary dimensions and $w_{m}$ and $w_{m'}$ are attention weights. This model has 16 free parameters.

We fit these models using different underlying 8D MDS solutions derived from human, VLM, CLIP and vision classifier data. Model fits were evaluated using the Bayesian Information Criterion (BIC) and percentage of variance accounted for ($\%Var$).

\section{Results}

\subsection{ Ratings and MDS Solutions}
\begin{figure}[h!]
\centering
\includegraphics[width=\linewidth]{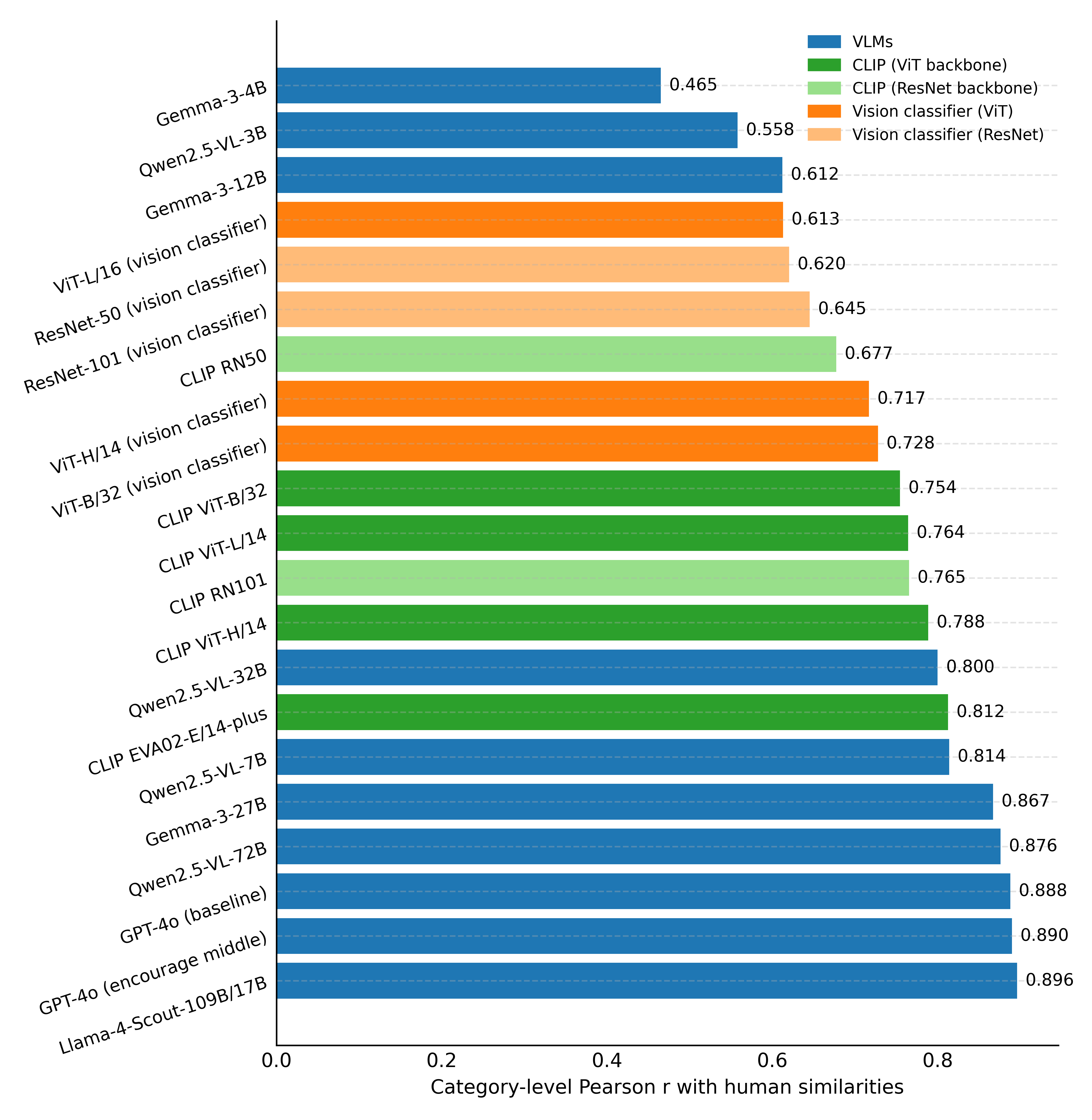}
\caption{Category-level Pearson correlations between human similarities and model-derived similarities.}
\label{fig:category_level_bars}
\end{figure}

\begin{figure}[ht]
    \centering
\includegraphics[width=0.85\textwidth]{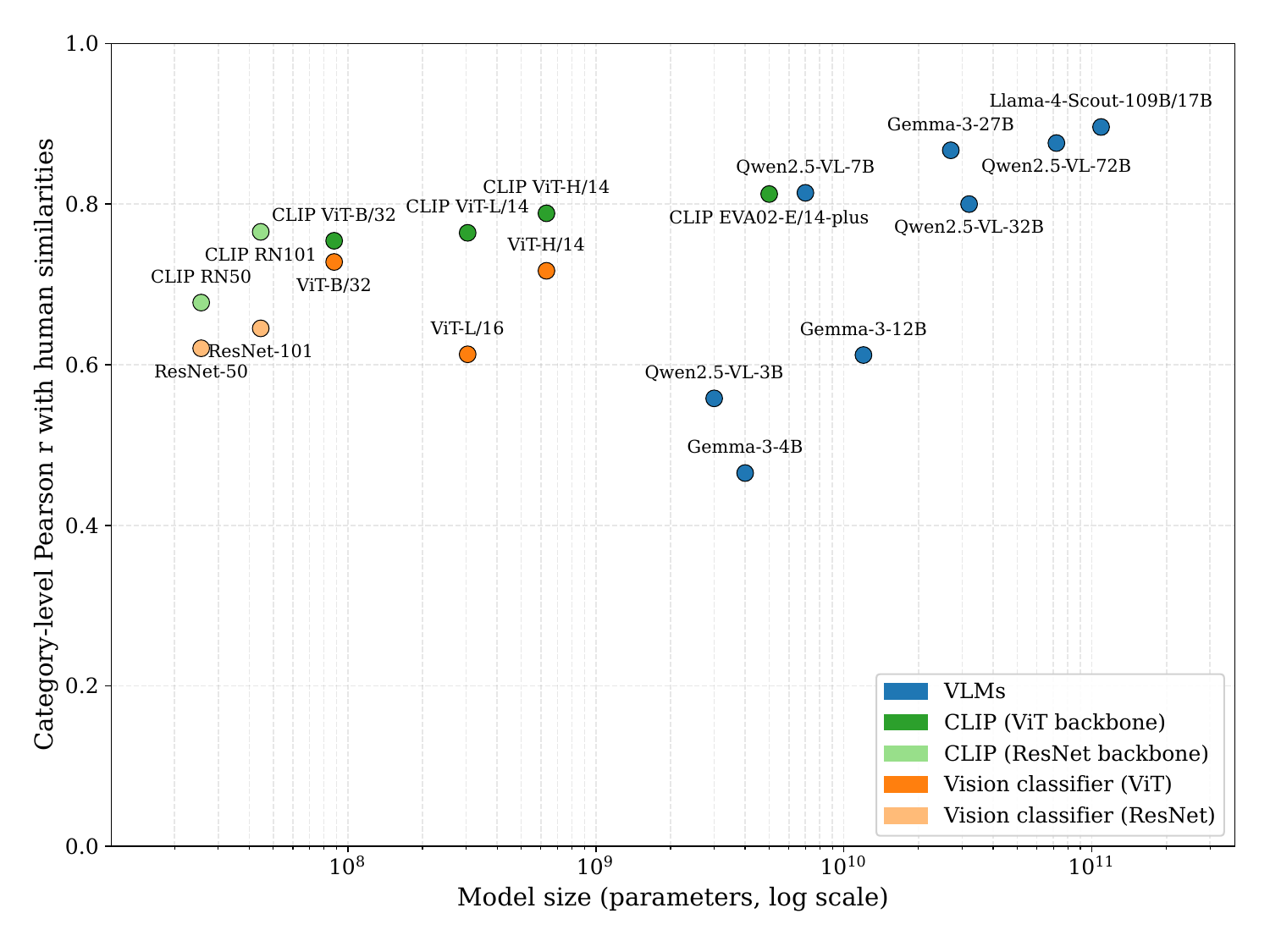}
    \caption{Model size versus human-model category-level correlation ($r$). GPT-4o is omitted because its parameter count has not been publicly disclosed.
For Llama-4-Scout (Mixture-of-Experts), we plot the total parameter count (109B) rather than the active parameter count used at inference (17B).}
    \label{fig:size_vs_corr_scatter}
\end{figure}

We first computed correlations between model- and human-derived \emph{category-level} similarity ratings (Figure~\ref{fig:category_level_bars}) and found strong alignment for most models. Correlations were highest for VLMs, followed by CLIP image encoders, and then vision-only classifiers. In general, larger models tended to yield higher correlations, with notable exceptions (Figure~\ref{fig:size_vs_corr_scatter}). We also computed correlations at the \emph{individual-pair} level and found them to be modest across all models (Figure~\ref{fig:individual_bars}), likely reflecting noise in the sparse human data (most pairs received only 1--2 human ratings).

\begin{figure}[ht]
\centering
\begin{subfigure}[t]{0.19\textwidth}
  \centering
  \includegraphics[width=\linewidth]{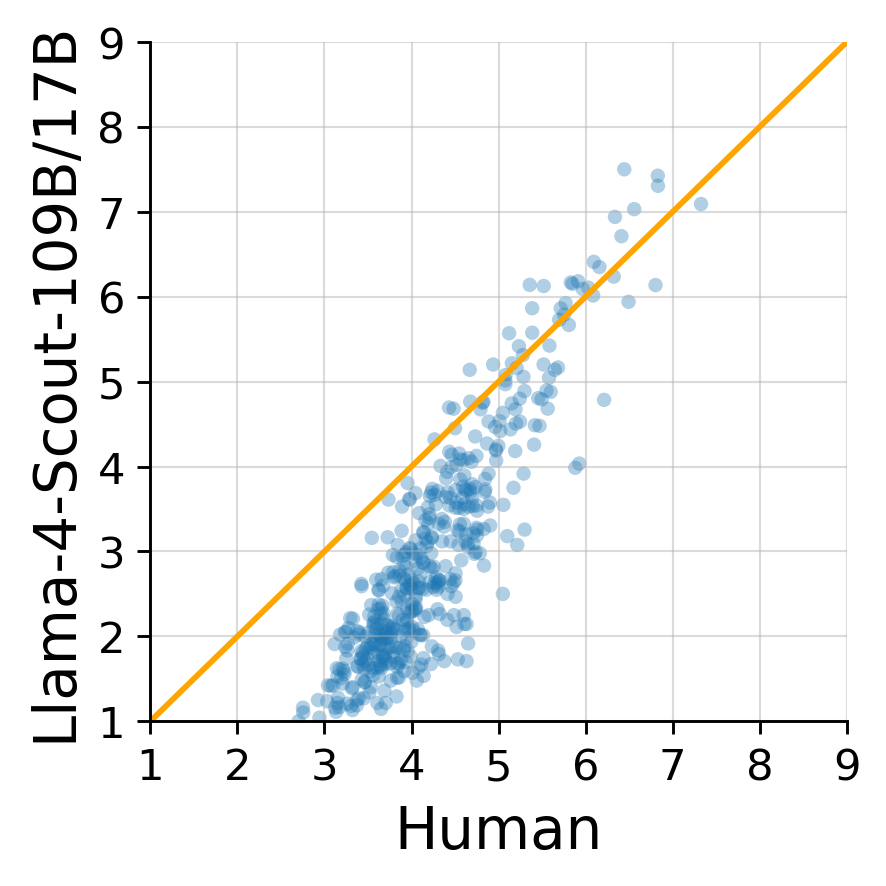}
\end{subfigure}\hfill
\begin{subfigure}[t]{0.19\textwidth}
  \centering
  \includegraphics[width=\linewidth]{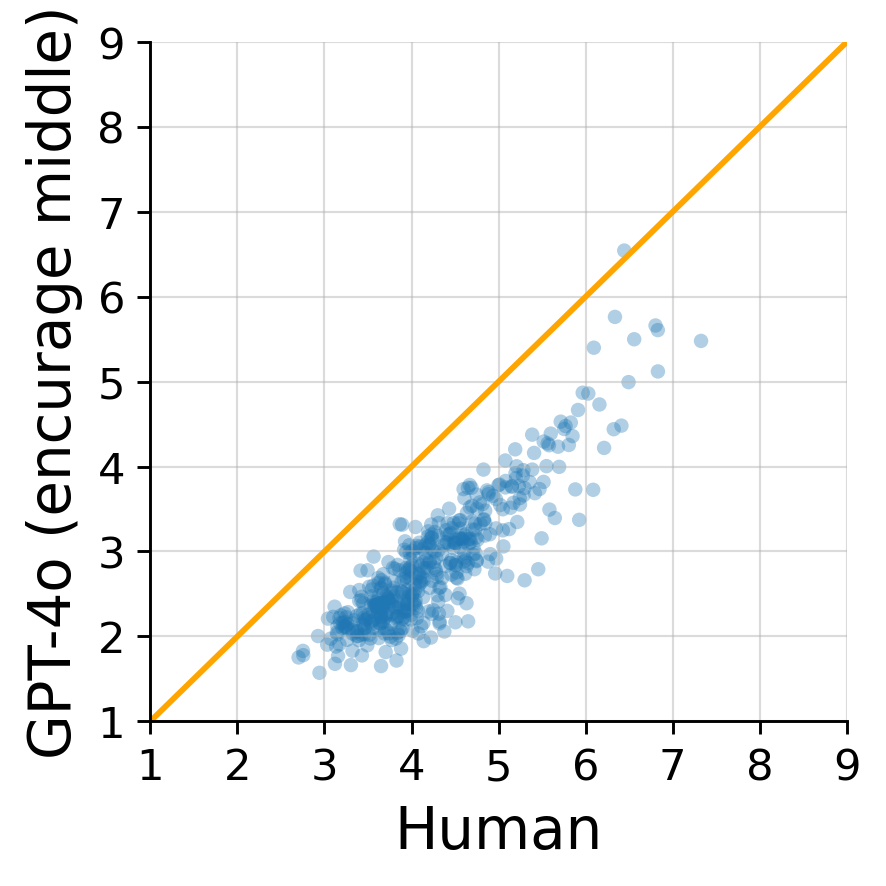}
\end{subfigure}\hfill
\begin{subfigure}[t]{0.19\textwidth}
  \centering
  \includegraphics[width=\linewidth]{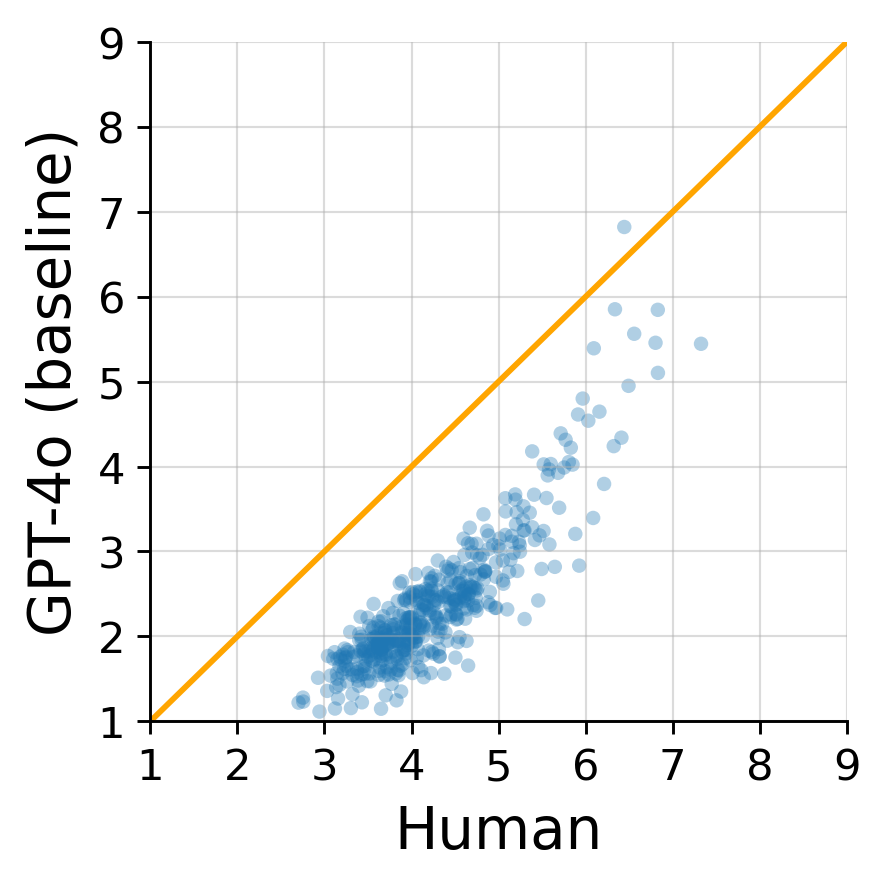}
\end{subfigure}\hfill
\begin{subfigure}[t]{0.19\textwidth}
  \centering
  \includegraphics[width=\linewidth]{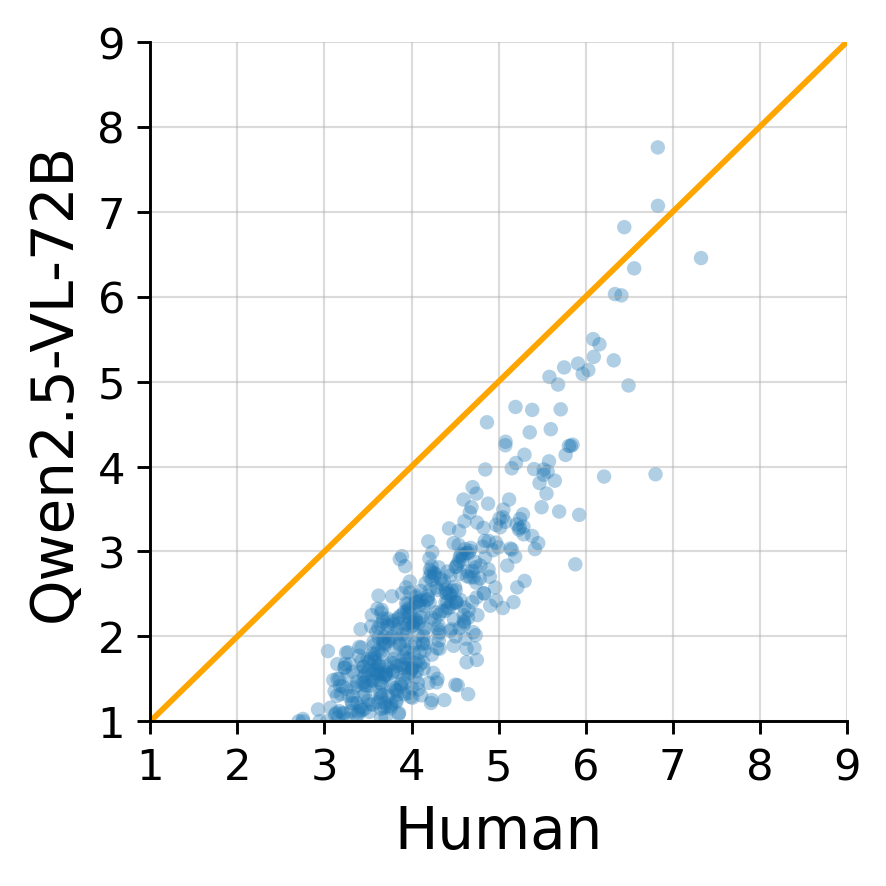}
\end{subfigure}\hfill
\begin{subfigure}[t]{0.19\textwidth}
  \centering
  \includegraphics[width=\linewidth]{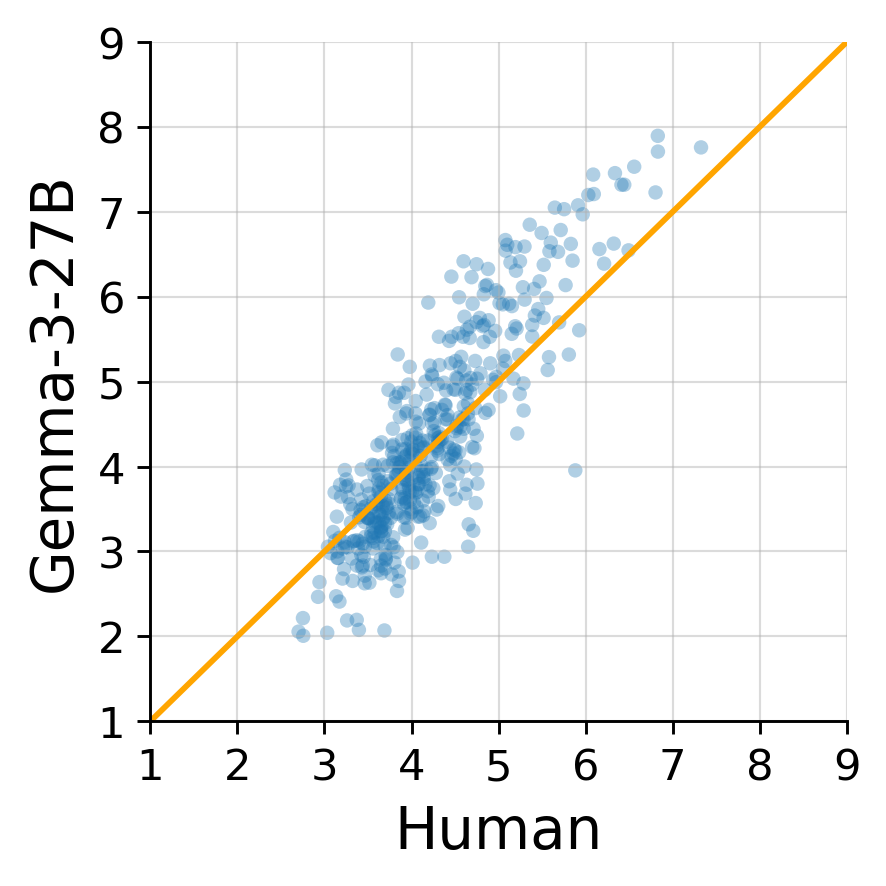}
\end{subfigure}

\vspace{0.6em}

\begin{subfigure}[t]{0.19\textwidth}
  \centering
  \includegraphics[width=\linewidth]{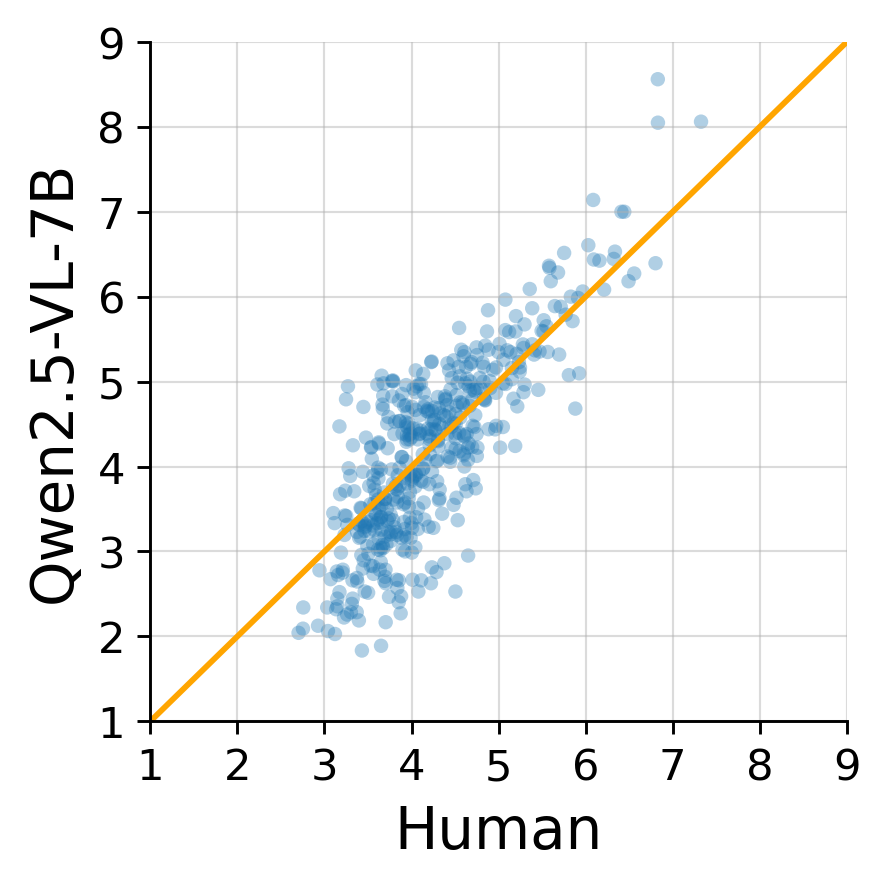}
\end{subfigure}\hfill
\begin{subfigure}[t]{0.19\textwidth}
  \centering
  \includegraphics[width=\linewidth]{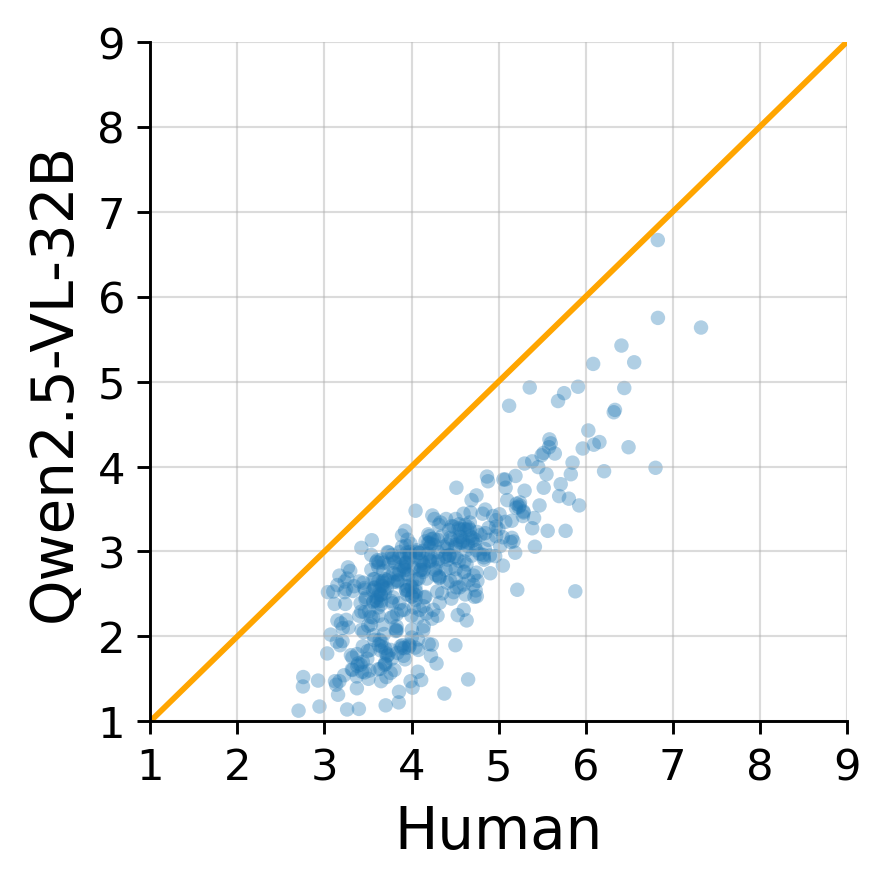}
\end{subfigure}\hfill
\begin{subfigure}[t]{0.19\textwidth}
  \centering
  \includegraphics[width=\linewidth]{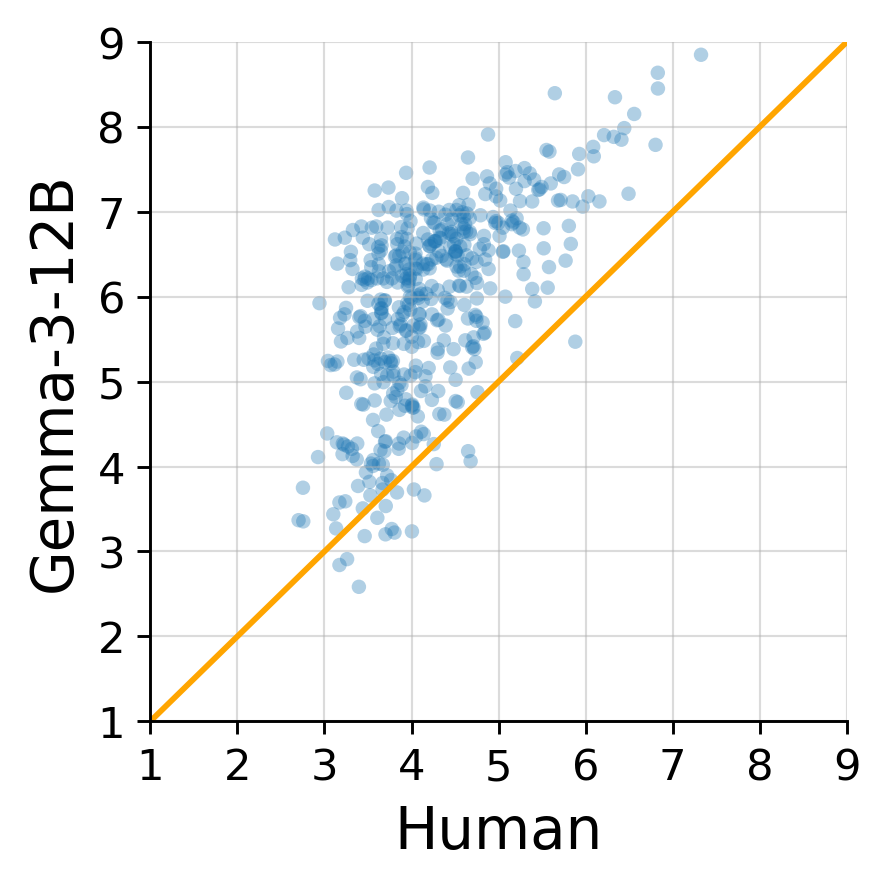}
\end{subfigure}\hfill
\begin{subfigure}[t]{0.19\textwidth}
  \centering
  \includegraphics[width=\linewidth]{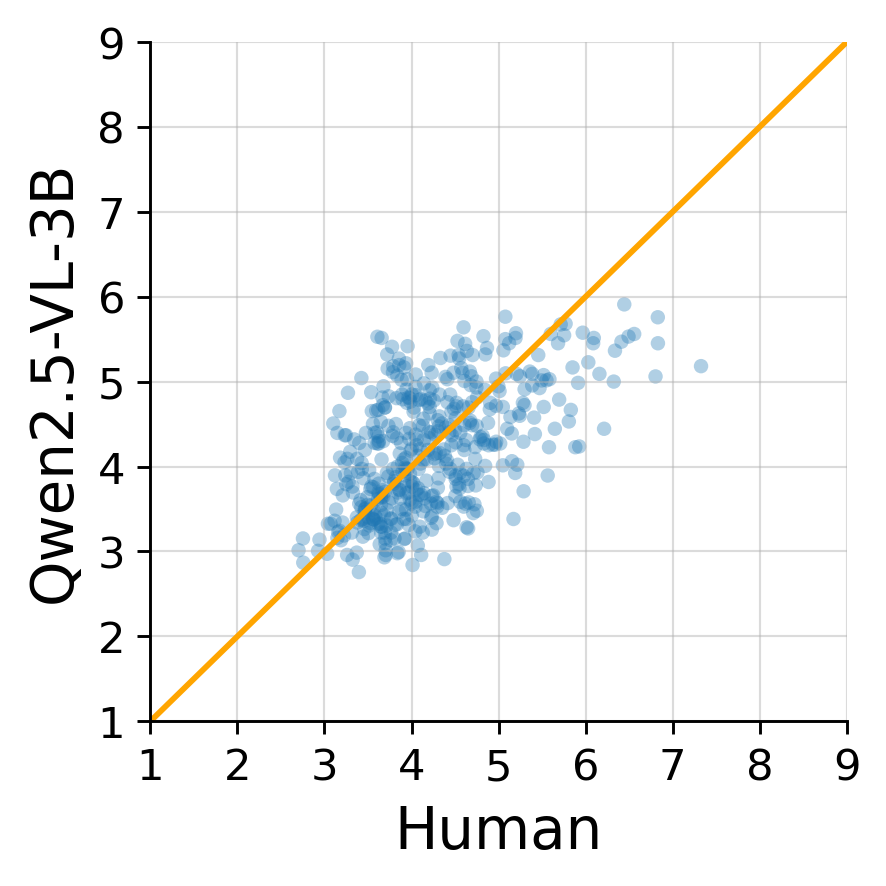}
\end{subfigure}\hfill
\begin{subfigure}[t]{0.19\textwidth}
  \centering
  \includegraphics[width=\linewidth]{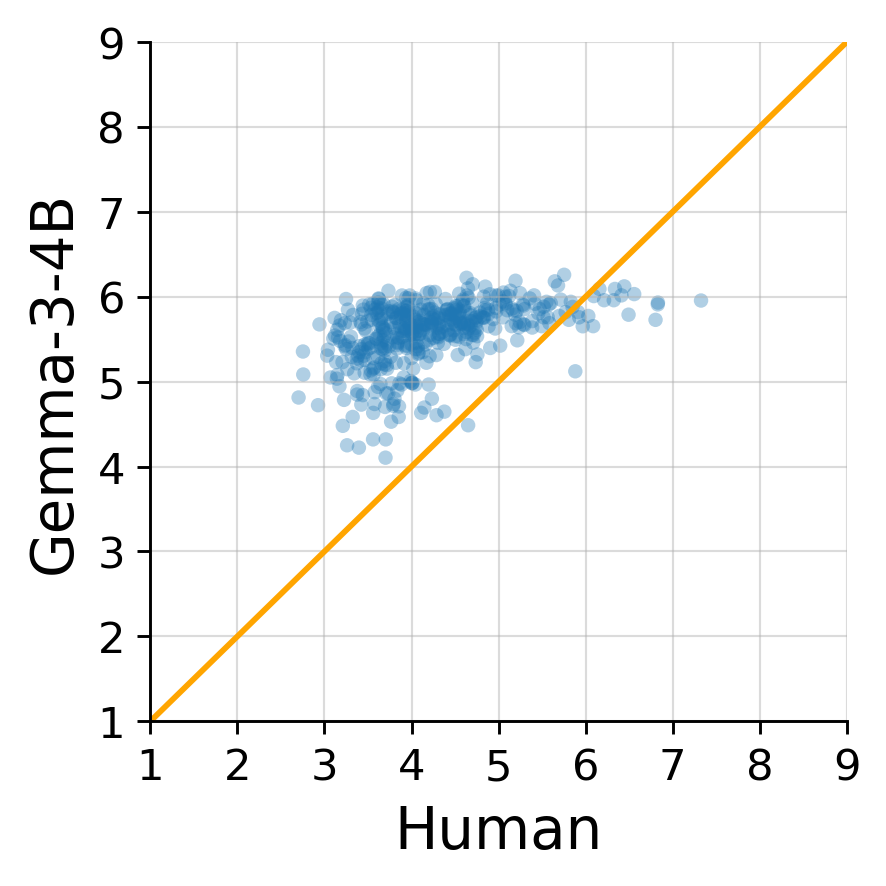}
\end{subfigure}

\caption{Category-level scatter plots (human vs.\ model) for VLMs. Points are category-pair entries; the orange line marks equality. Panels are ordered by Pearson correlation with human similarities (highest to lowest).}
\label{fig:vlm_category_scatter_grid}
\end{figure}

For VLMs, we provide scatterplots to visualize these relationships at both the category-level (Figure~\ref{fig:vlm_category_scatter_grid}) and the individual-pair level (Figure~\ref{fig:vlm_individual_scatter}). The category-level panels plot human mean similarities on the $x$-axis against model similarities on the $y$-axis for each of the $30 \times 29 / 2 = 435$ category pairs. These plots often revealed scale-use differences: for example, GPT-4o’s baseline prompt produces a distribution skewed toward the lower end of the 1-9 scale relative to humans (Figure~\ref{fig:rating_histograms}). An \textit{encourage middle} prompt broadened GPT-4o’s scale use, though both prompts remained distributionally distinct from humans. The ratings from the two GPT-4o prompts were very highly correlated with each other ($r=0.923$).

We focused subsequent analyses on a representative subset of models: four VLMs (GPT-4o, Gemma-3-27B, Qwen2.5-VL-72B, and Llama-4-Scout-109B/17B), one CLIP model (EVA02-E/14-plus), and one vision-only classifier (ResNet-50). 
For each model, we fit nonmetric MDS solutions at dimensionality 8 using Kruskal’s Stress-1 as the objective.  We focused our analyses on $p{=}8$ because, based on a combination of overall  fit and interpretability of the derived dimensions,  Nosofsky et al. (2018) had used an 8D solution for modeling the human similarity judgment data for these rock images. Setting $p{=}8$ for the solutions derived from the present machine learning models thereby enabled direct Procrustes alignment and straightforward comparisons across the human and model-derived MDS solutions. For GPT-4o we also report $p{=}10$ as an overcomplete embedding to test whether the additional degrees of freedom capture residual structure and improve alignment to the human 8D space. Because nonmetric MDS depends only on rank order, the two GPT-4o prompts yielded near-identical embeddings; all other models were analyzed with the baseline prompt only. The 8D (primary) solutions for each model are carried forward to the Procrustes and GCM analyses below.


\subsection{Correspondence with Human Psychological Space}
\begin{figure}[ht]
\centering
\setlength{\tabcolsep}{3pt}
\begin{tabular}{cccc}
\includegraphics[width=0.24\textwidth]{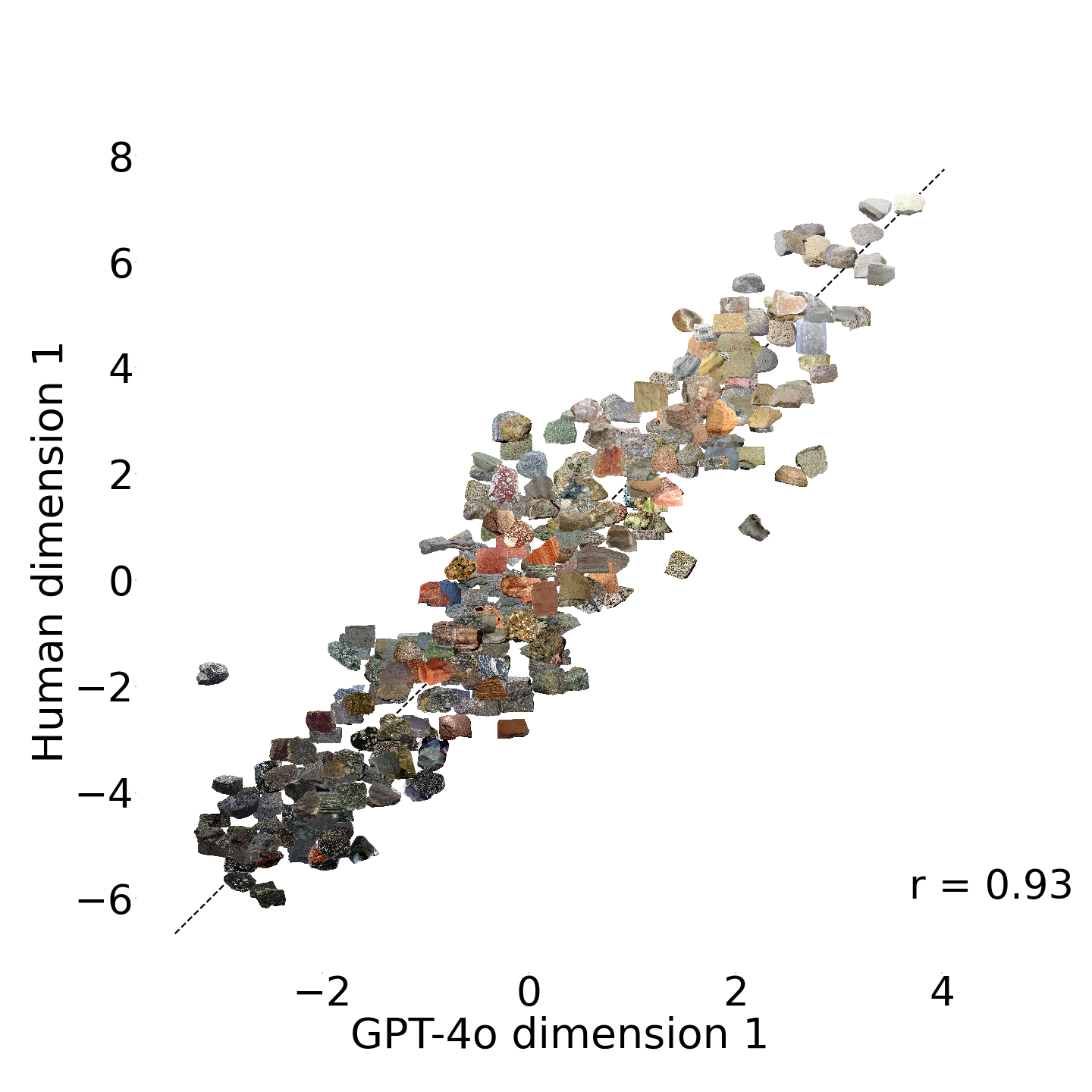} &
\includegraphics[width=0.24\textwidth]{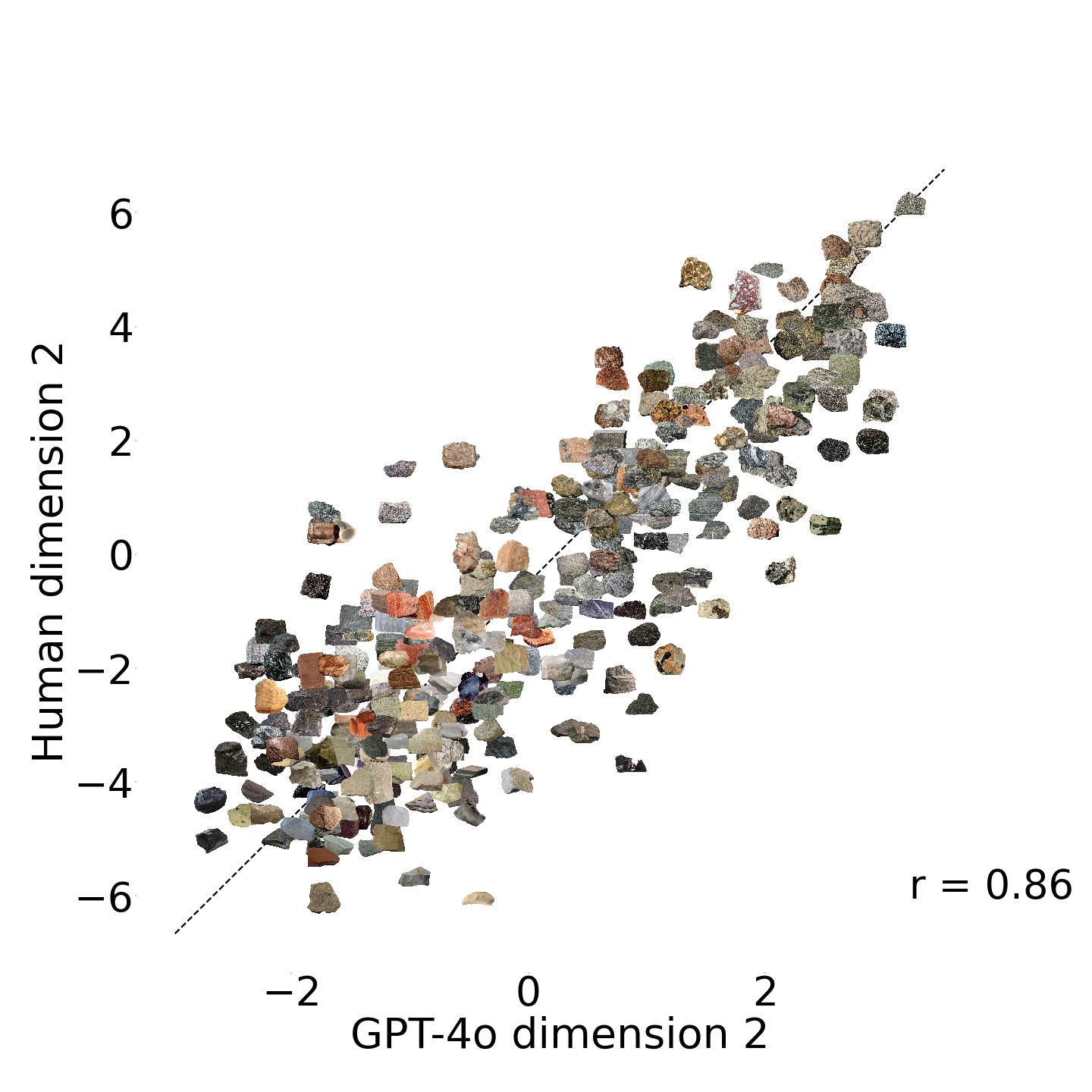} &
\includegraphics[width=0.24\textwidth]{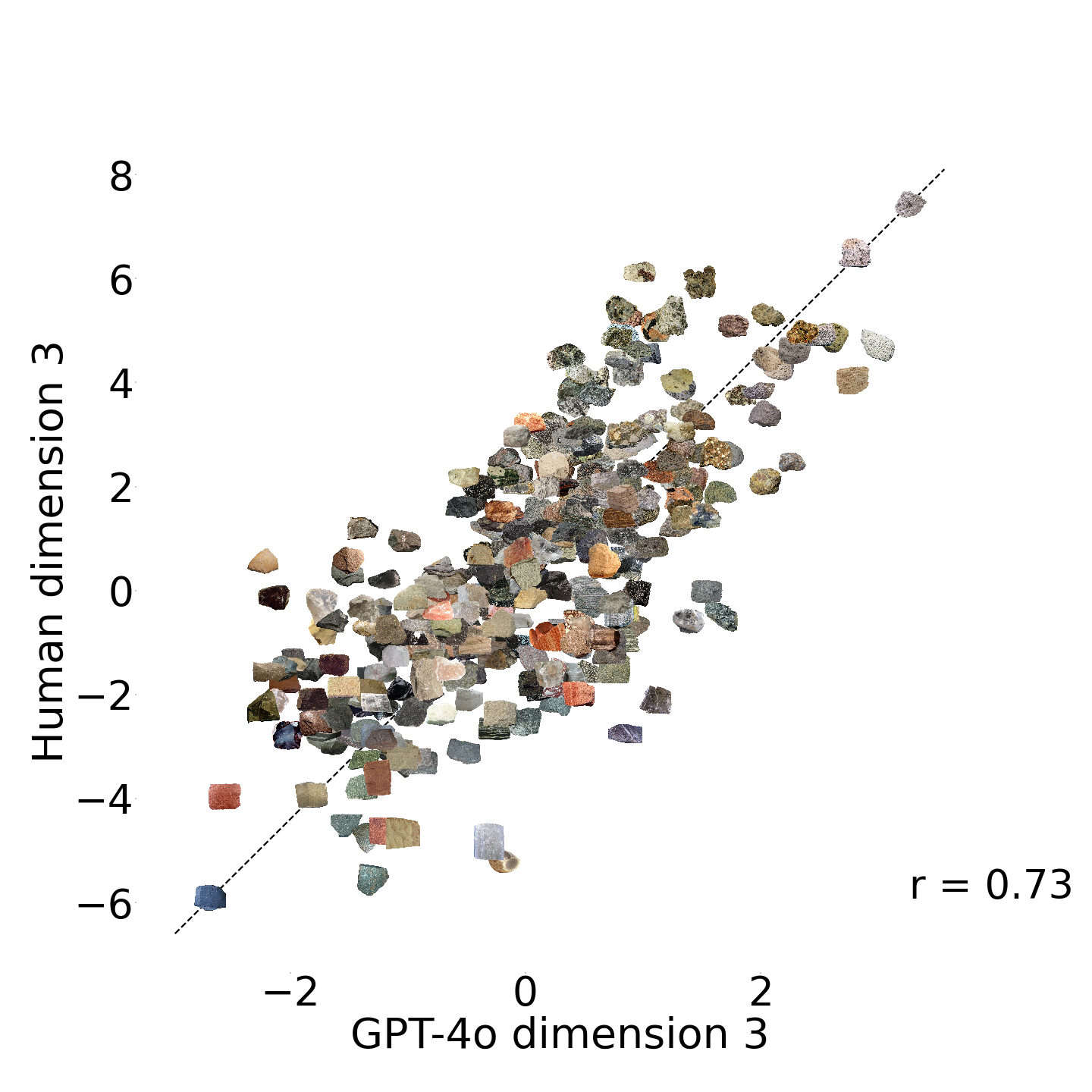} &
\includegraphics[width=0.24\textwidth]{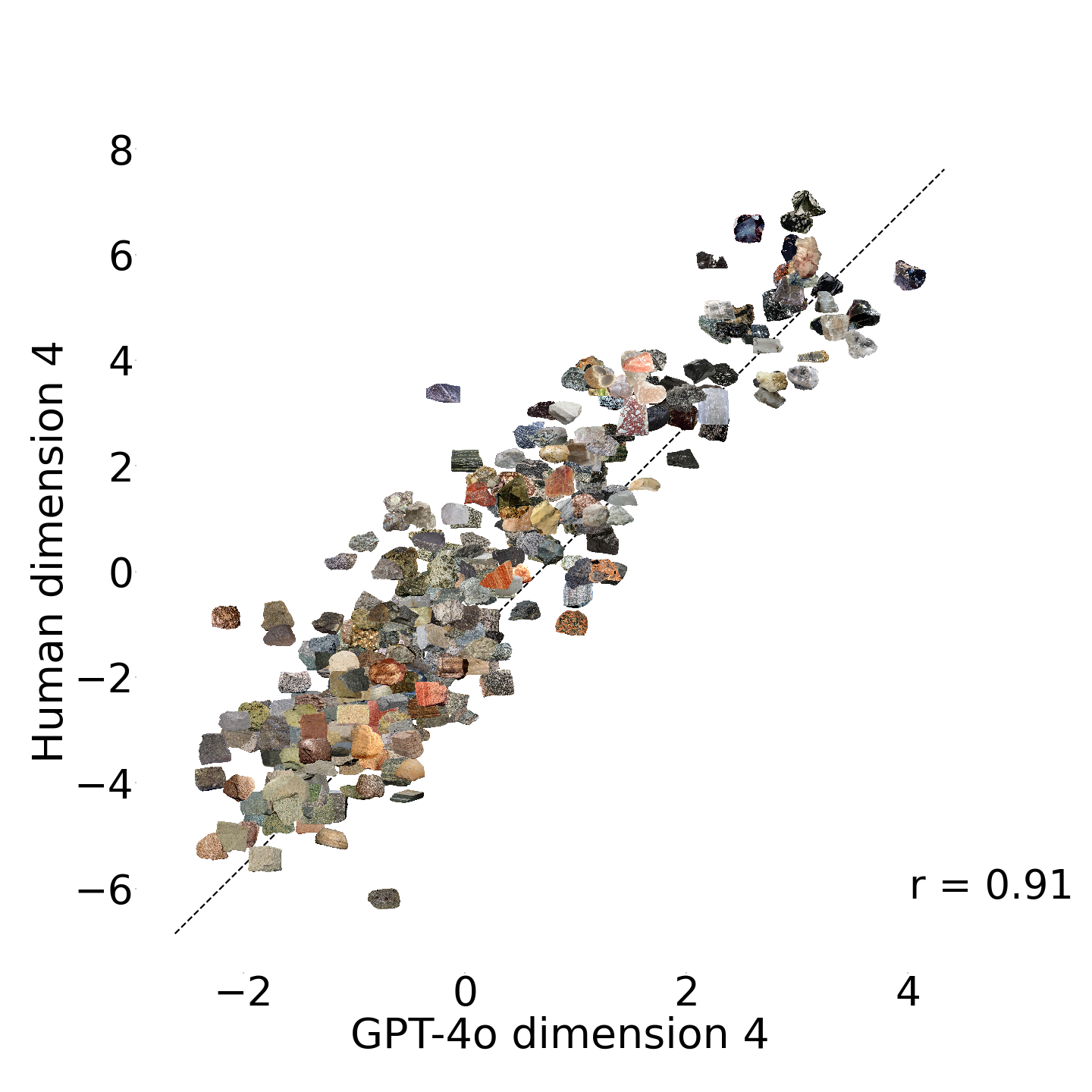} \\
\includegraphics[width=0.24\textwidth]{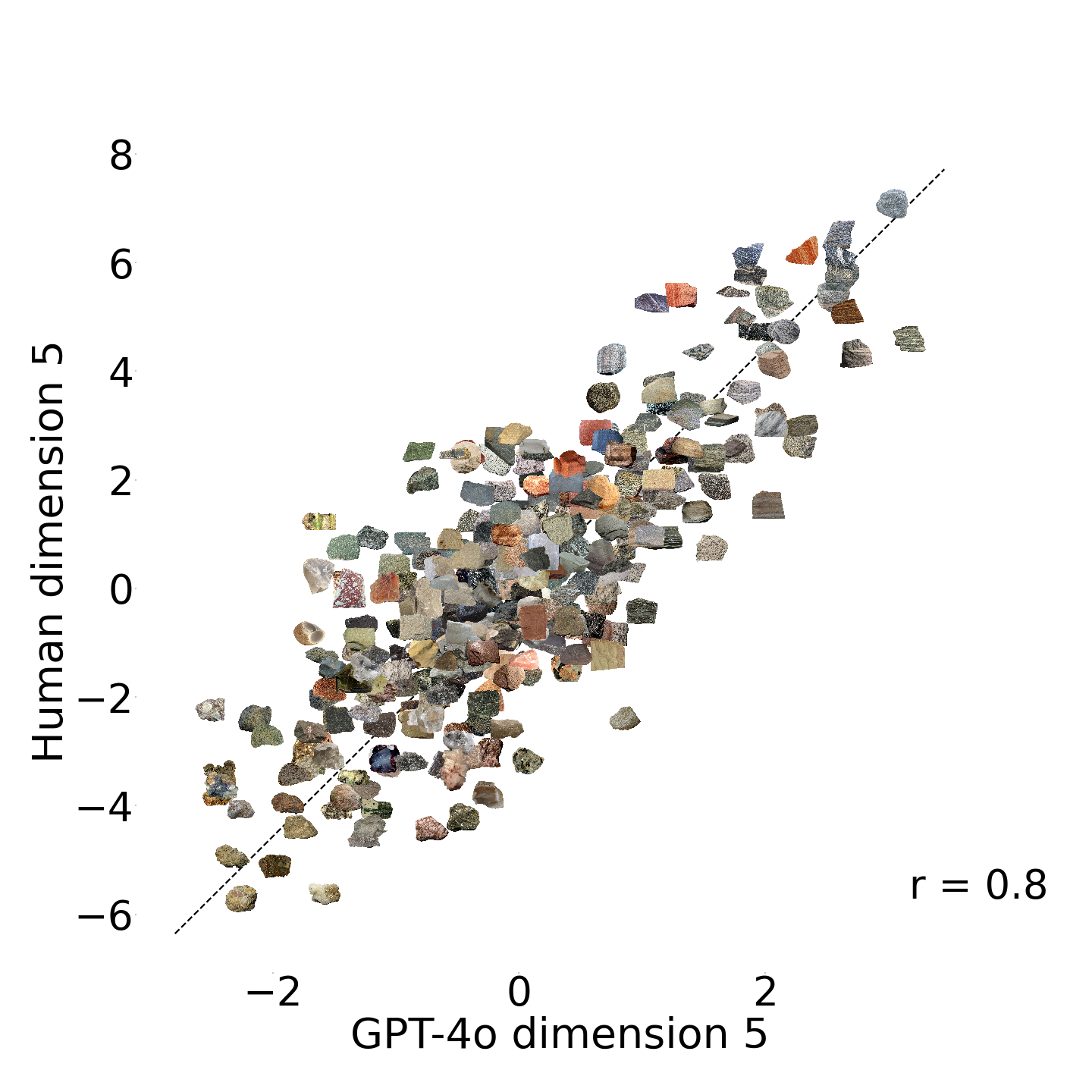} &
\includegraphics[width=0.24\textwidth]{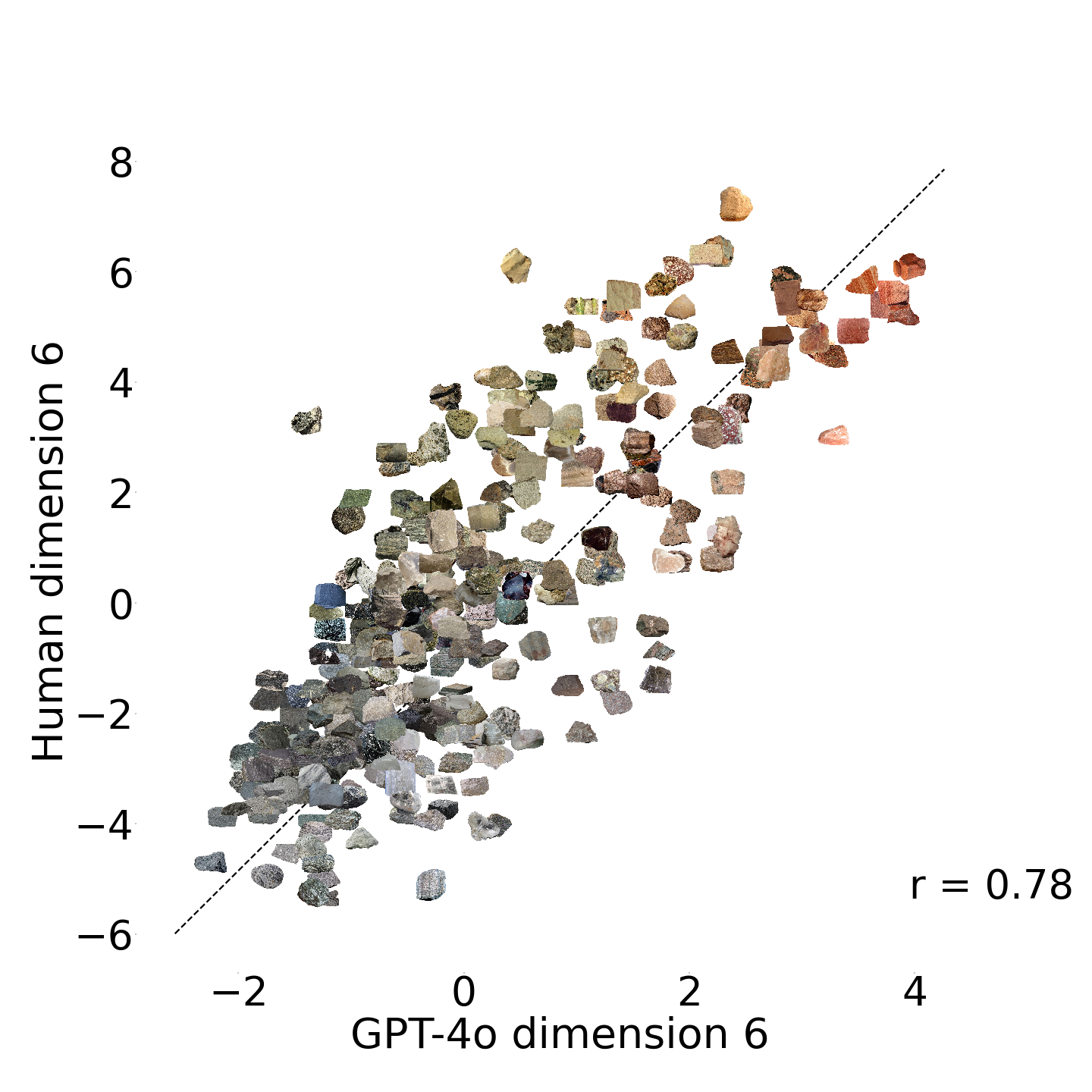} &
\includegraphics[width=0.24\textwidth]{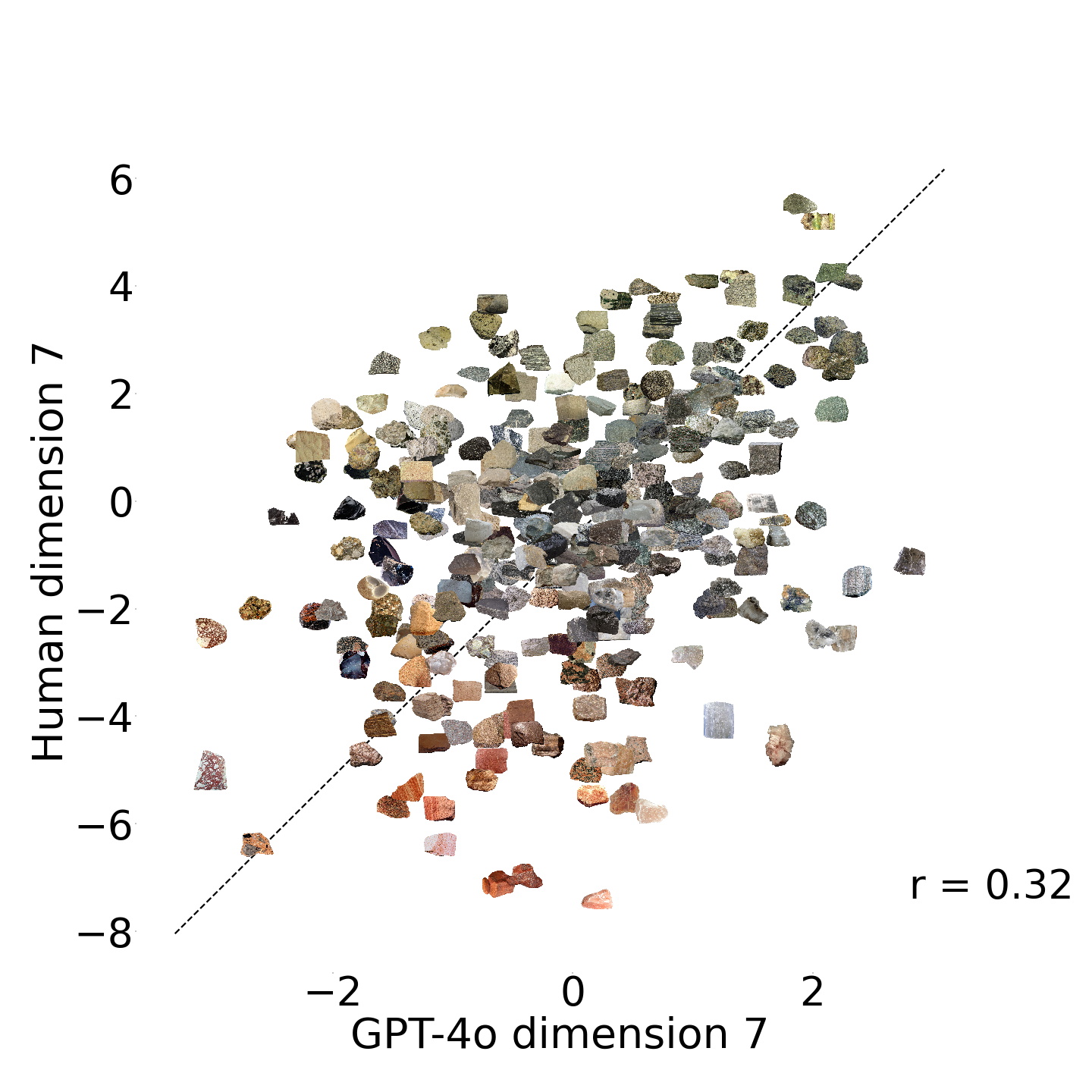} &
\includegraphics[width=0.24\textwidth]{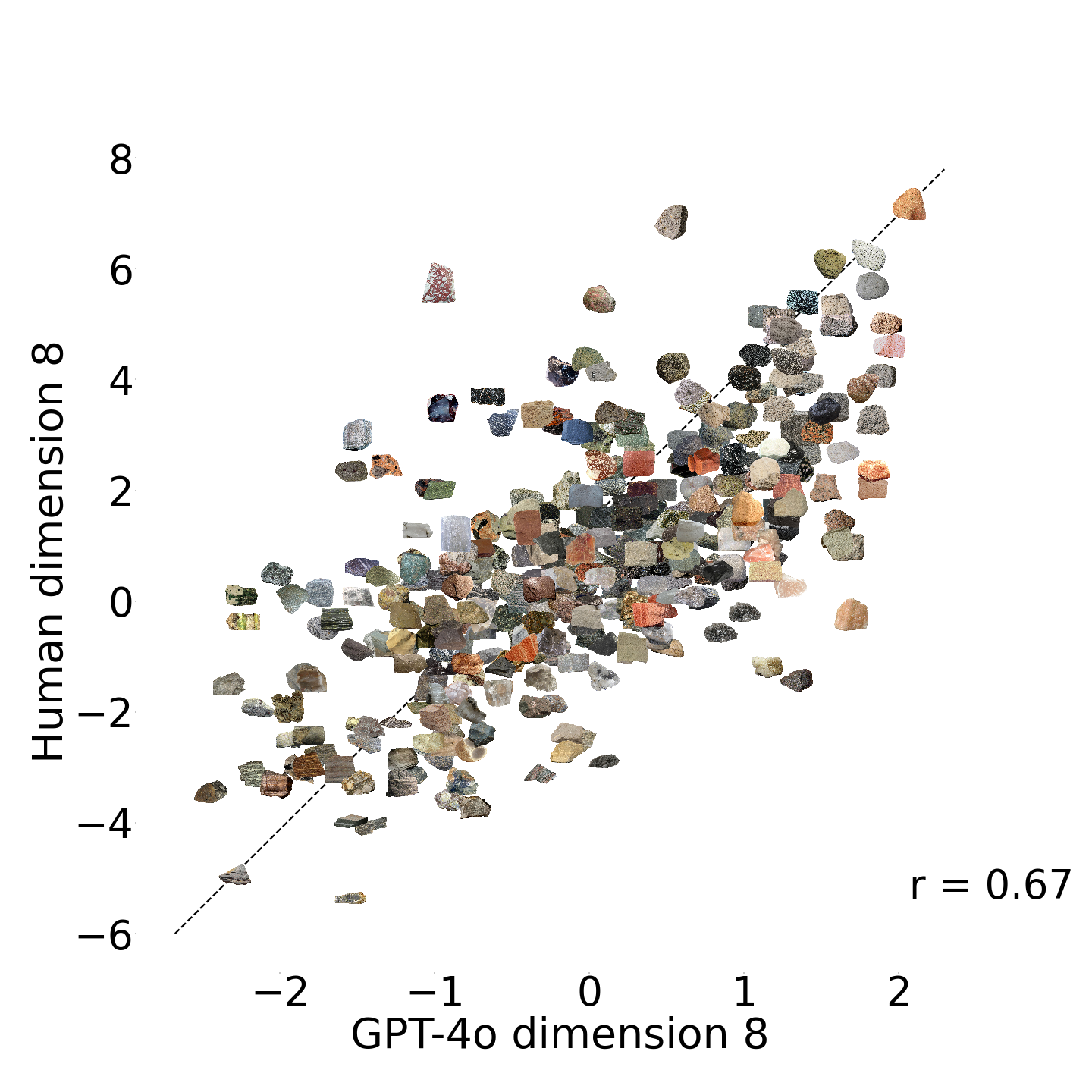} \\
\end{tabular}
\caption{Procrustes alignment between GPT-4o (encourage middle, 8D MDS) and the human 8D space, shown dimension by dimension. Each panel plots the aligned model coordinate (x-axis) against the corresponding human coordinate (y-axis) for all 360 rock images; the dashed line indicates equality and the in-panel $r$ is the Pearson correlation.}
\label{fig:gpt4o_procrustes_dims}
\end{figure}

We used Procrustes analysis to align each model’s solutions with the psychological dimensions derived from human data and report per-dimension Pearson correlations.
Table~\ref{tab:procrustes_8d_all} summarizes results for 8D solutions across models. Broadly, all families recover strong structure for lightness (D1), grain size (D2), shininess (D4), and moderate but highly significant correlations for D3 (rough/smooth), D5 (organization), D6 (chromaticity), and D8 (possibly shape-related). Hue (red/green; D7) is the most challenging: CLIP (EVA02-E/14-plus) and the  ResNet-50 classifier 
show weak alignment on D7 (0.09 and 0.009, respectively), whereas VLMs improve markedly (Qwen2.5-VL-72B: 0.75; Gemma-3-27B: 0.52; Llama-4-Scout-109B/17B: 0.41). GPT-4o exhibits high or moderately high correlations on all dimensions except for a lower correlation on D7 (0.32). The pattern is visualized in Figure~\ref{fig:gpt4o_procrustes_dims}, which plots the aligned GPT-4o coordinates against human coordinates for each dimension: D1, D2, D4, D5, and D6 show the strongest correlation, while D7 and D8 show noticeably greater scatter. Note that the high correlations arise because of good agreement across the entire continuous scale of dimension coordinates and not simply because of a few extreme values at the low and high ends of the scale.

For GPT-4o, moving from 8D to a 10D overcomplete solution and then rotating onto the human 8D space substantially improves alignment, especially on hue (D7 rises from 0.32 to 0.80; Table~\ref{tab:procrustes_10d_gpt} and Figure~\ref{fig:gpt4o_procrustes_dims_10}). This pattern suggests that human-relevant color/hue information is represented but distributed across additional degrees of freedom in the model space. In subsequent analyses, we carry forward the 8D solutions for all models (for comparability with the human space). 

\begin{table}[h!]
\centering
\caption{Procrustes correlations (Pearson $r$) between each model’s MDS dimensions and the human-rated dimensions.}
\label{tab:procrustes_8d_all}
\setlength{\tabcolsep}{4pt}
\small
\begin{tabular}{lccccccccc}
\toprule
\multicolumn{1}{c}{Model} &
\multicolumn{1}{c}{\rot{D1: Lightness}} &
\multicolumn{1}{c}{\rot{D2: Grain}} &
\multicolumn{1}{c}{\rot{D3: Texture}} &
\multicolumn{1}{c}{\rot{D4: Shiny}} &
\multicolumn{1}{c}{\rot{D5: Organization}} &
\multicolumn{1}{c}{\rot{D6: Chromaticity}} &
\multicolumn{1}{c}{\rot{D7: Hue (R/G)}} &
\multicolumn{1}{c}{\rot{D8}} &
\multicolumn{1}{c}{\rot{Mean}} \\
\midrule
\rowcolor{vlmblue!10}
Qwen2.5-VL-72B             & 0.919 & 0.818 & 0.564 & 0.860 & 0.717 & 0.833 & 0.745 & 0.621 & 0.760 \\
\rowcolor{vlmblue!10}
GPT-4o (encourage middle)     & 0.932 & 0.856 & 0.729 & 0.909 & 0.803 & 0.777 & 0.321 & 0.672 & 0.750 \\
\rowcolor{vlmblue!10}
GPT-4o (baseline)      & 0.930 & 0.857 & 0.697 & 0.912 & 0.795 & 0.770 & 0.320 & 0.684 & 0.746 \\
\rowcolor{vlmblue!10}
Llama-4-Scout-109B/17B            & 0.933 & 0.862 & 0.663 & 0.886 & 0.649 & 0.811 & 0.413 & 0.555 & 0.722 \\
\rowcolor{vlmblue!10}
Gemma-3-27B       & 0.912 & 0.842 & 0.558 & 0.771 & 0.612 & 0.755 & 0.522 & 0.547 & 0.690 \\
\addlinespace[2pt]
\rowcolor{clipgreen!10}
CLIP EVA02-E/14-plus   & 0.861 & 0.836 & 0.601 & 0.840 & 0.652 & 0.763 & 0.089 & 0.569 & 0.651 \\
\addlinespace[2pt]
\rowcolor{rnlight!25}
ResNet-50 (classifier) & 0.689 & 0.755 & 0.560 & 0.711 & 0.567 & 0.655 & 0.009 & 0.407 & 0.544\\
\bottomrule
\end{tabular}
\end{table}

\subsection{Predicting Human Classification}
We tested whether the MDS solutions derived from the models (VLM, CLIP or vision classifier) could substitute for human-derived solutions in predicting human classification behavior. We supply each model’s MDS embedding to GCM to predict human confusion probabilities from the classification experiment (see Section~\ref{sec:GCM} for more details about GCM and Section~\ref{sec:human_data} for more details about the human dataset). 

The results for the `core' GCM (using only the 8D MDS space) are shown in the left-hand side of Table \ref{tab:gcm_fits}. Strikingly, the models using the GPT-4o- and Llama-4-derived MDS solutions provided a substantially better fit to the human classification data than the model using the human-derived MDS solution, as indicated by lower BIC and a higher proportion of variance explained.

When we fit the extended GCM with 5 supplementary dimensions and attention weights attached to all dimensions (the right-hand side of Table \ref{tab:gcm_fits}), all models improved significantly. In this case, the GPT-4o-based models still performed better than the model using human MDS. The results were nearly identical for the `baseline' and `encourage middle' prompt MDS solutions. Aside from GPT-4o and Llama-4-Scout-109B/17B, other VLMs (Qwen2.5-72B and Gemma-3-27B) performed better than CLIP (EVA02-E/14-plus) and vision classifier (ResNet-50), but not as well as human MDS.  ResNet-50 was particularly low in the `core' GCM case, but still benefited from the five added dimensions. 

\begin{figure}[ht]
  \centering
  \includegraphics[width=\textwidth]{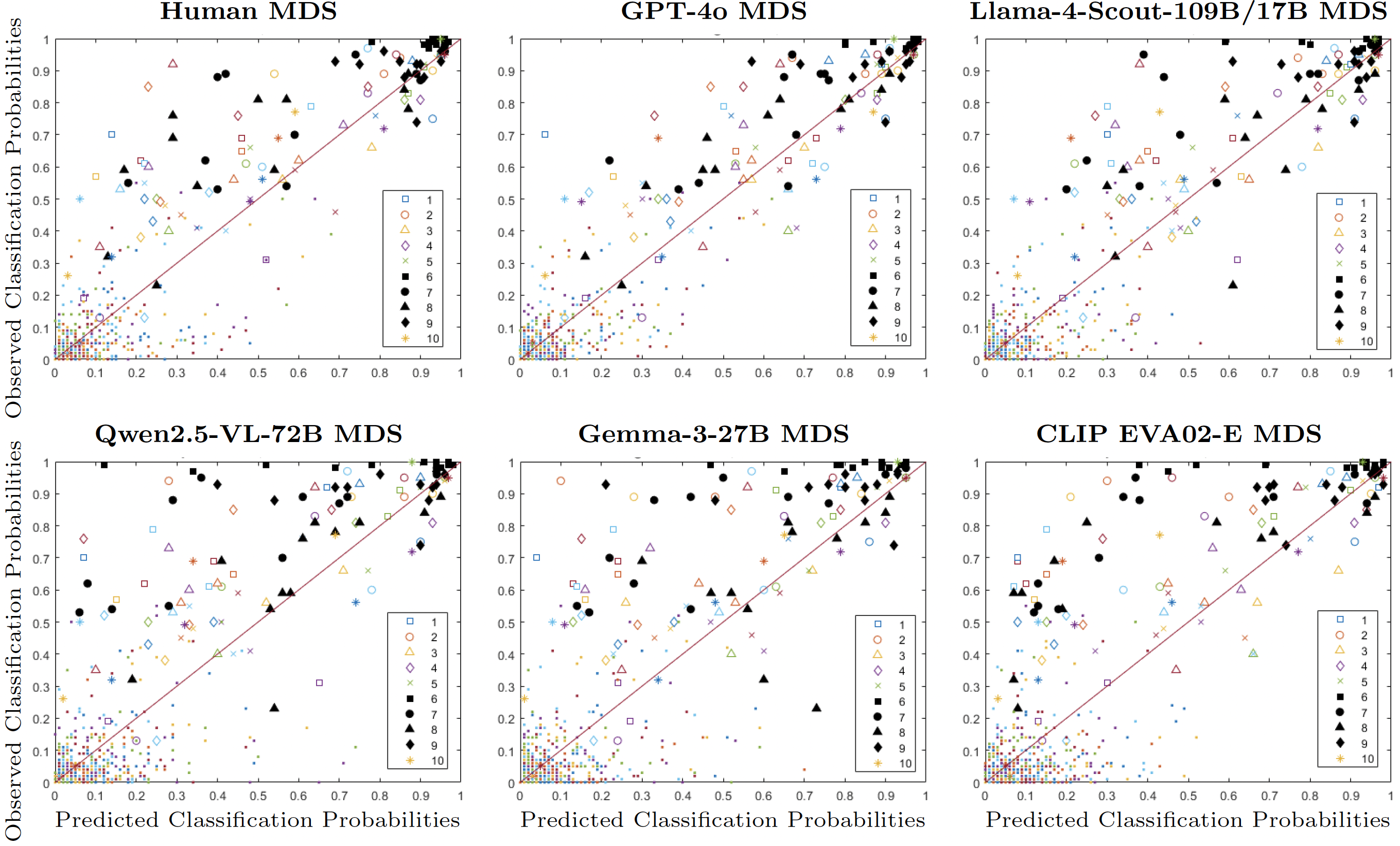}
  \caption{Predicted vs.\ observed classification probabilities for core GCMs.}
  \label{fig:gcm_scatterplots}
\end{figure}

Figures \ref{fig:gcm_scatterplots} and \ref{fig:gcm_scatterplots_supl}  show the scatterplots of observed versus model-predicted classification probabilities for a subset of models using `core' and `supplementary' dimensions respectively. Each panel plots, for every item--category cell in the $120\times10$ confusion matrix, the predicted classification probability against the observed human probability. Small dots denote off-diagonal cells (assignments to non-true categories). Shape markers denote the true-category cell for each of the 120 test items. Visual inspection confirms the quantitative results from Table \ref{tab:gcm_fits}, showing the superior fit of the supplementary-dimension models and the strong performance of the models based on the GPT-4o MDS solutions.

\begin{table}[h!]
\centering
\caption{GCM fits to the classification data. `Core' uses only the 8D MDS space; Supplementary augments with five rated dimensions. Lower BIC is better; higher \%Var is better. $P$ is the number of free parameters.}
\label{tab:gcm_fits}
\setlength{\tabcolsep}{5pt}
\small
\begin{tabular}{lcccccccc}
\toprule
\multirow{2}{*}{Model} &
\multicolumn{4}{c}{Core GCM (8D)} &
\multicolumn{4}{c}{Supplementary GCM (8D + 5 dims.)} \\
\cmidrule(lr){2-5} \cmidrule(lr){6-9}
& \rot{$P$} & \rot{$-\ln L$} & \rot{BIC} & \rot{\%Var} & \rot{$P$} & \rot{$-\ln L$} & \rot{BIC} & \rot{\%Var} \\
\midrule
\rowcolor{humangray}
Human MDS                    & 3  & 14559.2 & 29146.9 & 83.5 & 16 & 12896.7 & 25945.3 & 92.1 \\
\addlinespace[2pt]
\rowcolor{vlmblue!10}
GPT-4o (encourage middle)             & 3  & 13552.8 & 27134.3 & 89.5 & 16 & 12789.0 & 25730.0 & 92.8 \\
\rowcolor{vlmblue!10}
GPT-4o (baseline)                     & 3  & 13621.2 & 27271.0 & 89.1 & 16 & 12801.7 & 25755.3 & 92.7 \\
\rowcolor{vlmblue!10}
Llama-4-Scout-109B/17B                           & 3  & 14096.4 & 28221.3 & 86.0 & 16 & 12980.6 & 26113.2 & 91.8 \\
\rowcolor{vlmblue!10}
Qwen2.5-VL-72B                                 & 3  & 15590.0 & 31208.5 & 75.7 & 16 & 13280.5 & 26713.0 & 89.4 \\
\rowcolor{vlmblue!10}
Gemma-3-27B                      & 3  & 16029.8 & 32088.1 & 74.4 & 16 & 13416.3 & 26984.6 & 89.4 \\
\addlinespace[2pt]
\rowcolor{clipgreen!10}
CLIP (EVA02-E/14-plus)                & 3  & 16381.5 & 32791.5 & 72.9 & 16 & 13690.3 & 27532.6 & 89.0 \\
\addlinespace[2pt]
\rowcolor{rnlight!25}
ResNet-50 (vision classifier)         & 3  & 20436.3 & 40901.1 & 55.0 & 16 & 14149.9 & 28451.7 & 84.0 \\
\bottomrule
\end{tabular}
\end{table}


\section{Discussion}

This study demonstrates that psychological spaces derived from VLMs can show strong geometric alignment with human perceptual dimensions and serve as a superior substrate for predicting human categorization behavior compared to spaces derived from human judgments themselves. Specifically, in a naturalistic domain with expert-grounded structure, spaces recovered from VLM pairwise judgments align with human-derived axes under simple Procrustes transforms, and when supplied to the GCM they exceed the predictive utility of spaces constructed from human similarity ratings. These results advance beyond global human-model agreement by identifying the axes themselves and by demonstrating that those axes can help explain human behavior. Beyond suggesting shared constraints on perceptual organization across human and VLM learning systems, our findings have enormous practical implications: VLMs can be used to closely simulate human similarity judgments. The collection of such judgments from humans has been essential for deriving the types of feature spaces that serve as inputs to a wide variety of computational models of cognitive processes. This human data collection process becomes infeasible as the size of the relevant stimulus set becomes large \cite{Goldstone1994,Hout2013,Richie2020,Hebart2023}. Thus, by making use of VLMs to provide the similarity judgments, the bottleneck is removed, allowing for more widespread application of computational models to large-scale cognitive studies.

Across model families, we observe a performance hierarchy: VLM-raters, which make explicit pairwise judgments, produce psychological spaces that better predict human categorization than fixed-embedding models like CLIP, which in turn outperform standard vision classifiers. A mechanistic explanation for this may lie in the emergent computational capabilities of the Transformer architecture. Recent work suggests that Vision Transformers can spontaneously develop a two-stage processing architecture: a ``perceptual stage'' for extracting object features, followed by a ``relational stage'' for explicitly comparing them \cite{lepori2024beyond}. Our fixed-embedding approach, which calculates distances between independently generated vectors, likely taps only the perceptual stage. In contrast, the pairwise similarity prompt forces the VLM to engage its latent relational stage, performing a joint computation over both inputs that better captures the human comparative judgments.

We found that the VLM-derived space, when used as input to a classic exemplar model (GCM), predicts human classification better than a space derived from human similarity data. We propose this occurs because VLM judgments act as a denoised surrogate for human perception. Human pairwise ratings are notoriously noisy, affected by fatigue, inter-rater variability, and sparse sampling \cite{Goldstone1994,Hout2013}. The VLM, trained on a vast corpus of visual-textual data, may capture a more robust, idealized form of the underlying perceptual structure. This hypothesis is supported by large-scale replications of psychology experiments where LLMs produced cleaner data and larger effect sizes than the original human studies \cite{cui2024can}, and by findings that humans, when unaware of an artwork's origin, systematically prefer AI-generated art, suggesting models can capture a ``supernormal'' version of human aesthetic structure \cite{vanhees2025human}. The VLM-derived space may therefore be a more effective input for the GCM precisely because it represents a less noisy, idealized version of the perceptual geometry that is only imperfectly measured from human participants.

Our axis-level finding complements and refines prior alignment work. Global geometry comparisons including Representational Similarity Analysis (RSA) \cite{RSA2008,Haxby2014,Nili2014Toolbox} and network-similarity indices such as Centered Kernel Alignment (CKA) and Projection-Weighted Canonical Correlation Analysis (PWCCA) \cite{Kornblith2019CKA,Morcos2018PWCCA} establish broad correspondences across brains and models. By rotating model-derived MDS spaces onto human normative axes, we ask which interpretable factors align (e.g., lightness, grain, surface properties) rather than only whether two high-dimensional geometries correlate. This axis-level readout can be used alongside RSA/CKA/PWCCA, together with judgment-space metrics (e.g., Turing RSA \cite{ogg2025turingRSA}), to provide complementary views: overall similarity of geometries, stability of internal features, and the identification of psychologically meaningful factors.

At the same time, alignment of geometry is not a mechanistic explanation. LLMs/VLMs are not faithful simulators of human cognition, can be prompt- or wording-sensitive, and reflect training-distribution biases  \cite{stureborg2024llmjudge,schroder2025large,birhane2021multimodal}. High geometric correspondence does not guarantee equivalence of features or computations \cite{Diedrichsen2017,Walther2016}, and shortcut solutions remain a pervasive risk \cite{Geirhos2020}. We therefore view our findings through a ``proxy'' lens: models need not be theories of mind to provide representational fuel for classic cognitive theories \cite{ziv2025large}. In this light, the success of VLM-derived spaces strengthens the explanatory reach of exemplar-based accounts when supplied with rich, semantically informed coordinates.

Our approach is complementary to other modern methods for constructing psychological spaces, such as Deep Metric Learning (DML), which learns a direct mapping from stimuli to a psychological space by fitting human behavioral data \cite{Sanders2020,leonvillagra2024learning}, and to work that pairs psychological embeddings with radial basis function networks to predict exemplar- and category-level ease of human category learning from similarity data \cite{RoadsMozer2021RBF}. While DML and RBFN-based approaches reduce the data burden compared to classic MDS, our VLM-rater method eliminates the need for new human data collection and yields interpretable spaces. Interestingly, these seemingly competing approaches show signs of theoretical convergence. Researchers in related fields are increasingly using language to guide and regularize visual representation learning, for example, by using LMM-generated semantic descriptions to improve CLIP embeddings for classification \cite{tzelepi2024lmm}. This suggests a broader paradigm shift: moving from deriving psychological structure from purely perceptual data towards a new approach that recognizes the fundamental role of language and semantics in shaping human visual representation, a move compatible with classic geometric accounts of concepts \cite{Gardenfors2000,Aka2023Memorability}.

The present work is focused on a single, albeit complex, naturalistic domain. Future work should establish the generalizability of these findings to other domains and tasks. A particularly exciting direction lies in opening the black box of the VLM's decision process. While Procrustes analysis confirms that the VLM's psychological dimensions align with human-rated ones, it does not confirm that they are grounded in the same low-level visual features. Techniques from Explainable AI (XAI) that produce interpretable visualizations of which image regions contribute to a model's similarity judgment could be applied to directly test whether the VLM and humans are ``looking'' at the same evidence (e.g., grain size, texture patterns) when making their judgments \cite{black2022visualizing,Chefer2021TransformerExplain,Selvaraju2017GradCAM}. Such hybrid approaches, combining scalable measurement from foundation models with rigorous cognitive theory and new tools for interpretability, can help advance a theory-grounded science of the mind that is equipped to handle the complexity of the real world \cite{Kerrigan2021ConfusionCalibration}.

In sum, identifying \emph{dimension-level convergence} and \emph{behavioral sufficiency} brings foundation model analysis closer to classic theories of psychological space \cite{Shepard1987,Tversky1977}. Practically, model-elicited pairwise judgments plus ordinal embeddings provide a scalable path to psychologically meaningful coordinates; scientifically, the observed convergence points to shared constraints on perceptual organization across distinct learning systems. Integrating axis-level alignment with neural measurements (e.g., RSA/CKA across brain, behavior, and model spaces) and with \emph{active} pairwise sampling \cite{Jamieson2011,Vankadara2023} will help chart when and why such convergence arises.
As cognitive modeling and machine learning continue to converge, such hybrid approaches can help scale rigorous, theory-grounded analyses to the complexity of real-world stimuli, advancing both scientific understanding and practical methodology.

\bibliography{references}


\renewcommand{\thetable}{A\arabic{table}}
\renewcommand{\thefigure}{A\arabic{figure}}
\setcounter{table}{0}
\setcounter{figure}{0}

\clearpage

\section*{Supplementary Information}

\begin{figure}[h!]
    \centering
    \includegraphics[width=\textwidth]{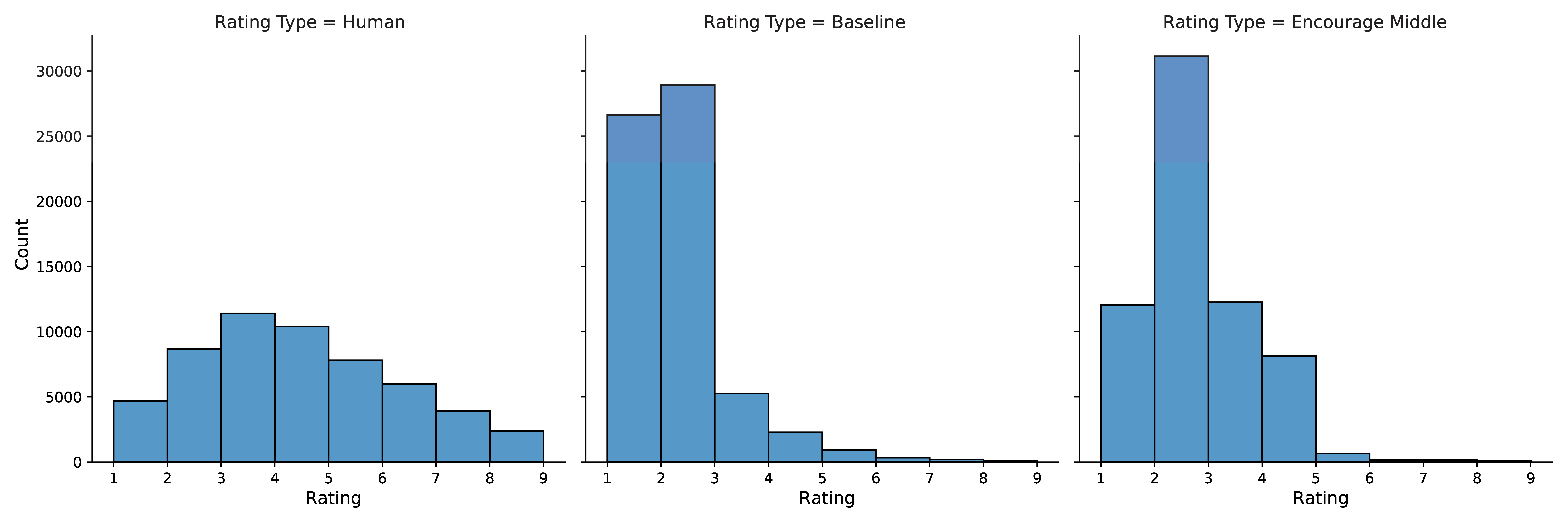}
    \caption{Distribution of similarity ratings for the rocks dataset. (Left) Human ratings. (Middle) GPT-4o ratings using the `baseline' prompt. (Right) GPT-4o ratings using the `encourage middle' prompt.}
    \label{fig:rating_histograms}
\end{figure}

\begin{figure}[h!]
\centering
\includegraphics[width=\linewidth]{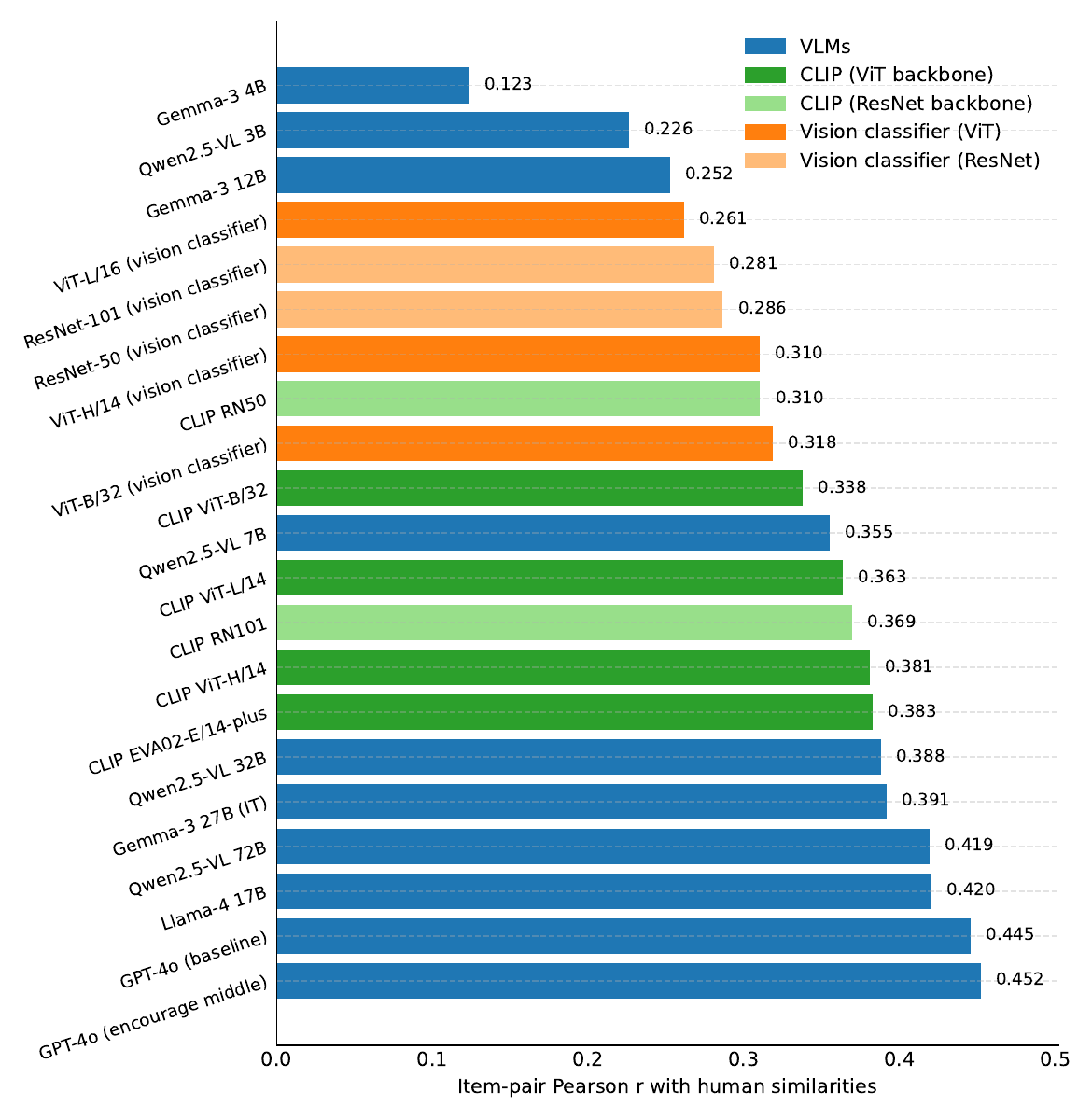}
\caption{Item-level Pearson correlations between human similarities and model-derived similarities.}
\label{fig:individual_bars}
\end{figure}

\begin{figure}[ht]
\centering
\begin{subfigure}[t]{0.19\textwidth}
  \centering
  \includegraphics[width=\linewidth]{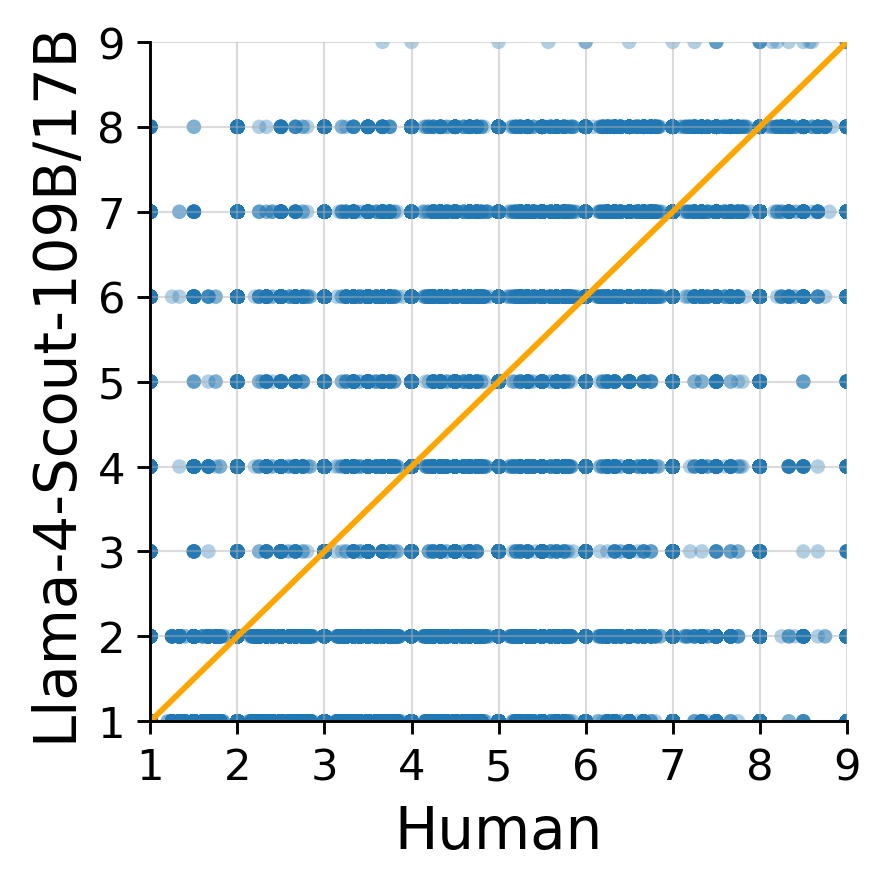}
\end{subfigure}\hfill
\begin{subfigure}[t]{0.19\textwidth}
  \centering
  \includegraphics[width=\linewidth]{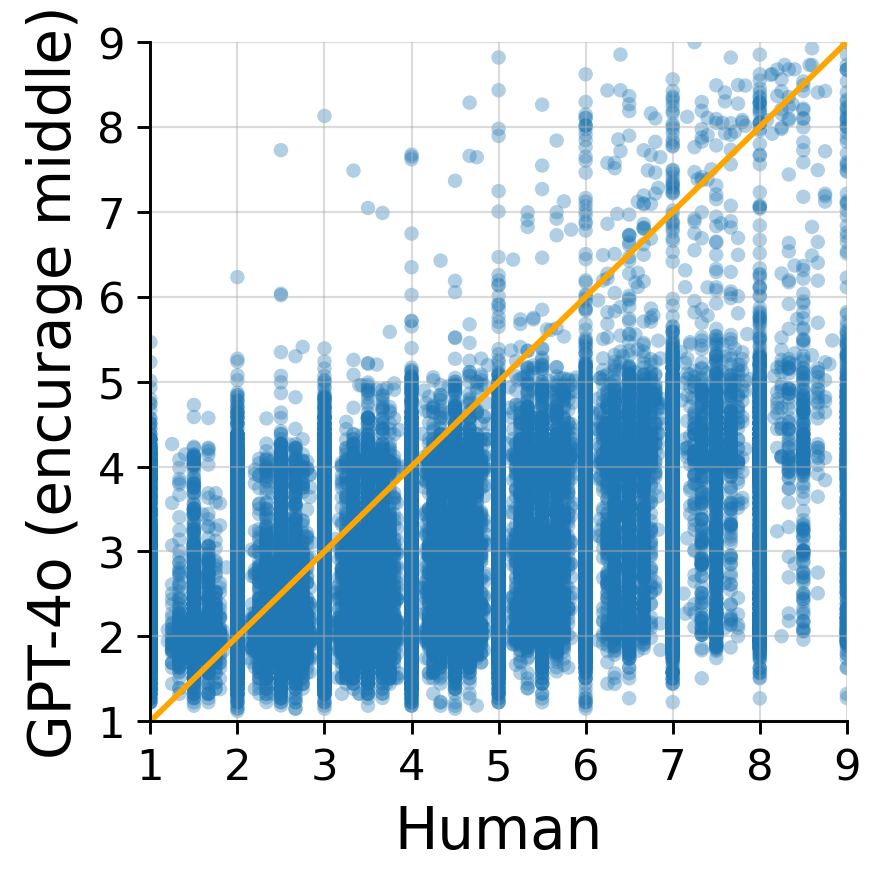}
\end{subfigure}\hfill
\begin{subfigure}[t]{0.19\textwidth}
  \centering
  \includegraphics[width=\linewidth]{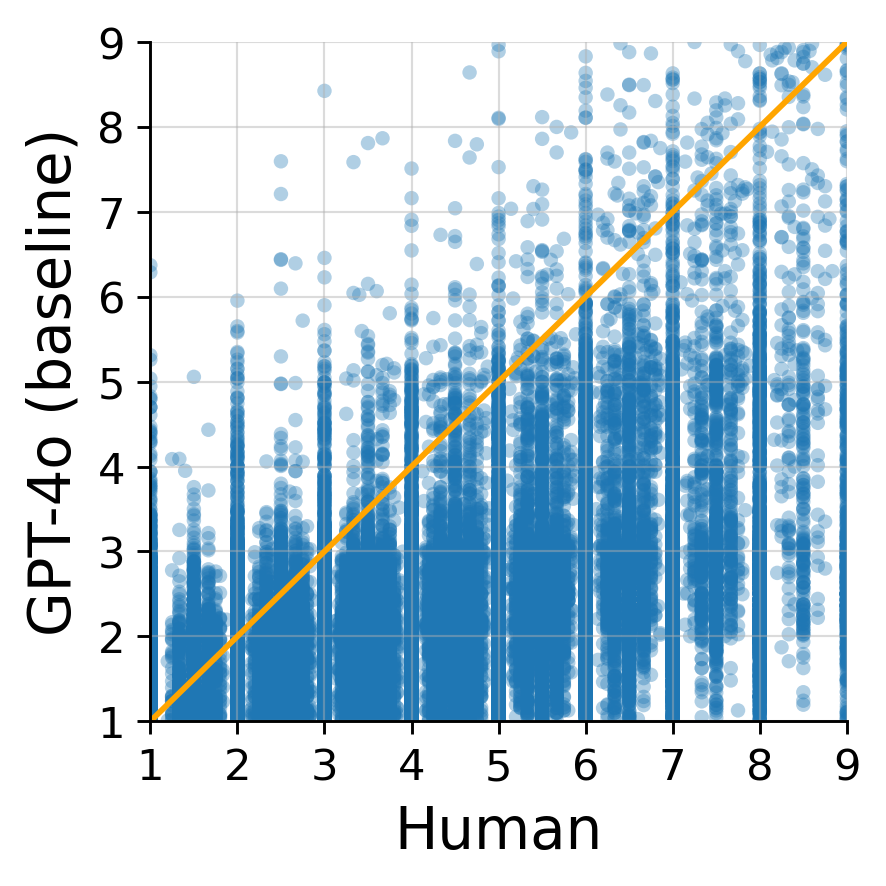}
\end{subfigure}\hfill
\begin{subfigure}[t]{0.19\textwidth}
  \centering
  \includegraphics[width=\linewidth]{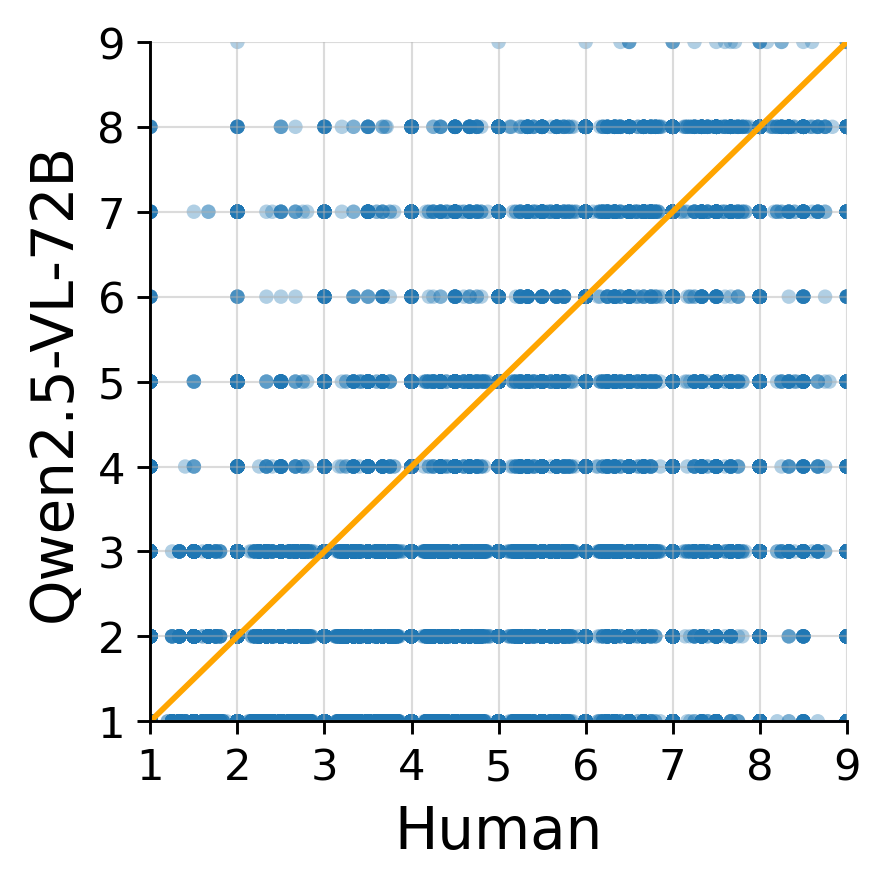}
\end{subfigure}\hfill
\begin{subfigure}[t]{0.19\textwidth}
  \centering
  \includegraphics[width=\linewidth]{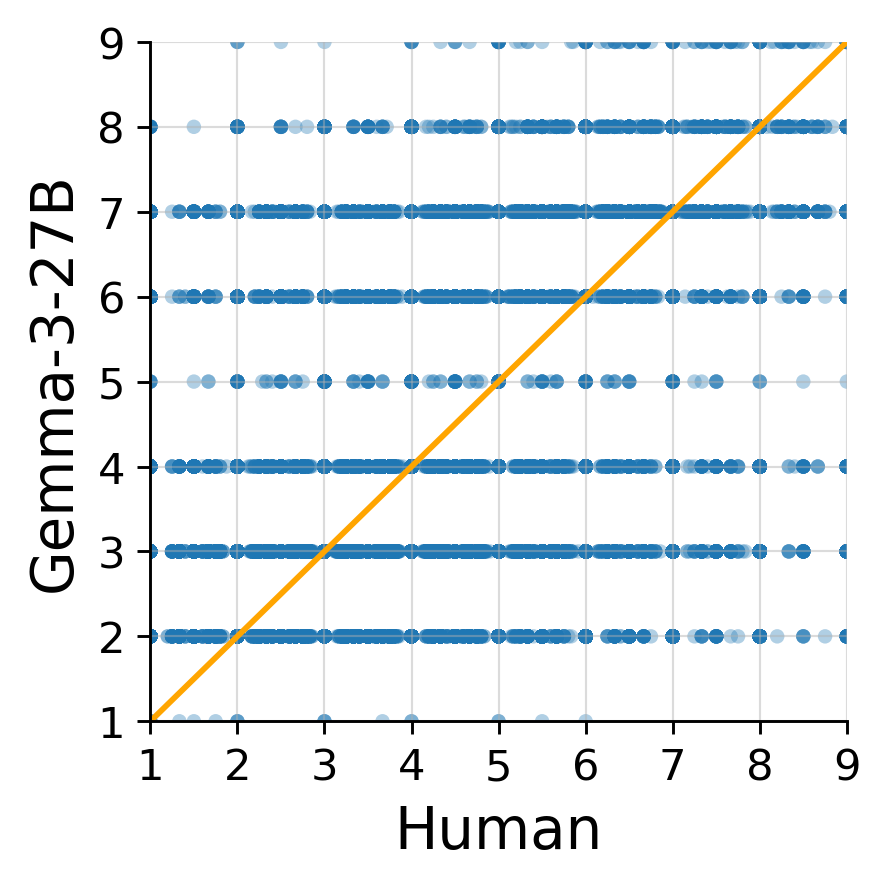}
\end{subfigure}

\vspace{0.6em}

\begin{subfigure}[t]{0.19\textwidth}
  \centering
  \includegraphics[width=\linewidth]{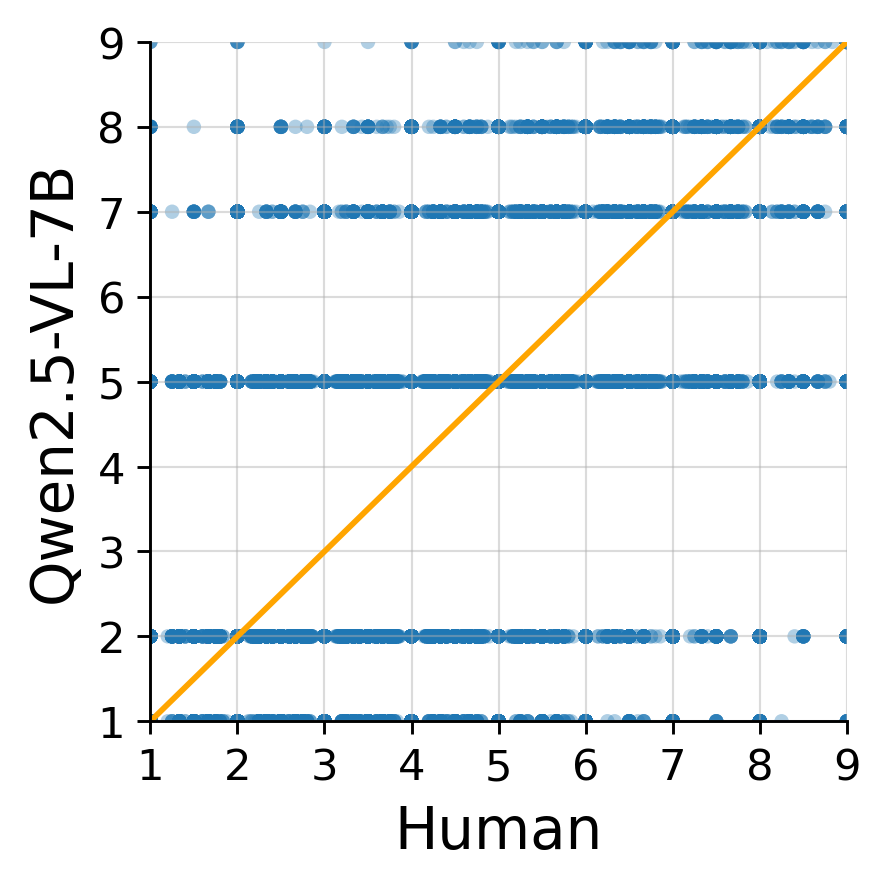}
\end{subfigure}\hfill
\begin{subfigure}[t]{0.19\textwidth}
  \centering
  \includegraphics[width=\linewidth]{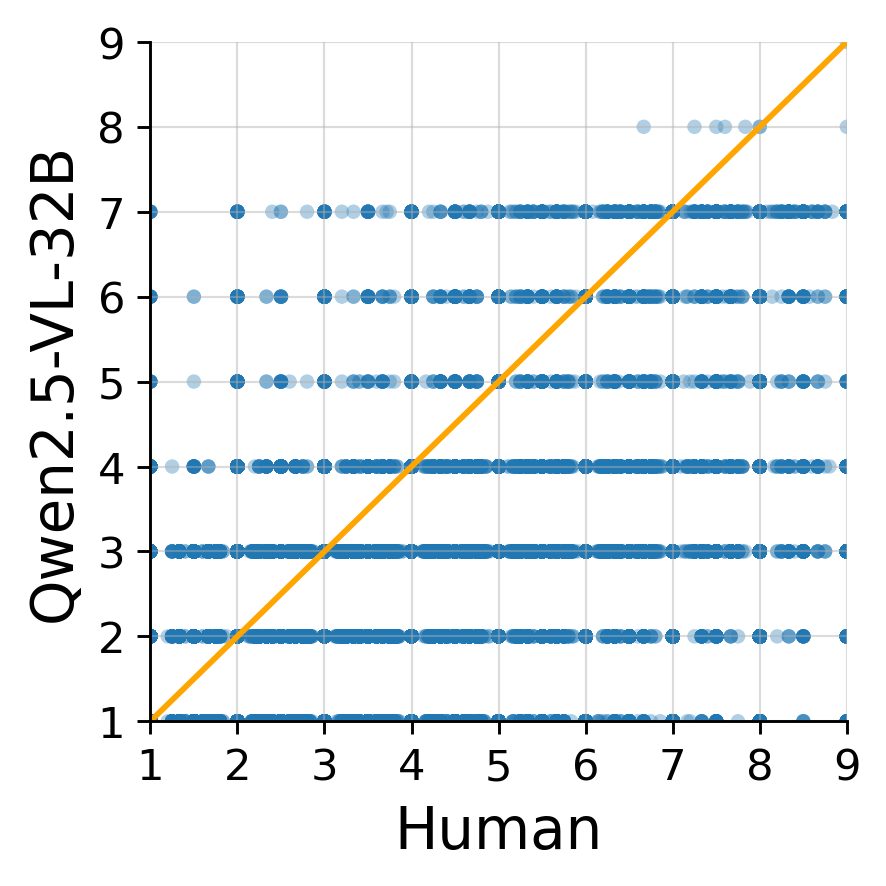}
\end{subfigure}\hfill
\begin{subfigure}[t]{0.19\textwidth}
  \centering
  \includegraphics[width=\linewidth]{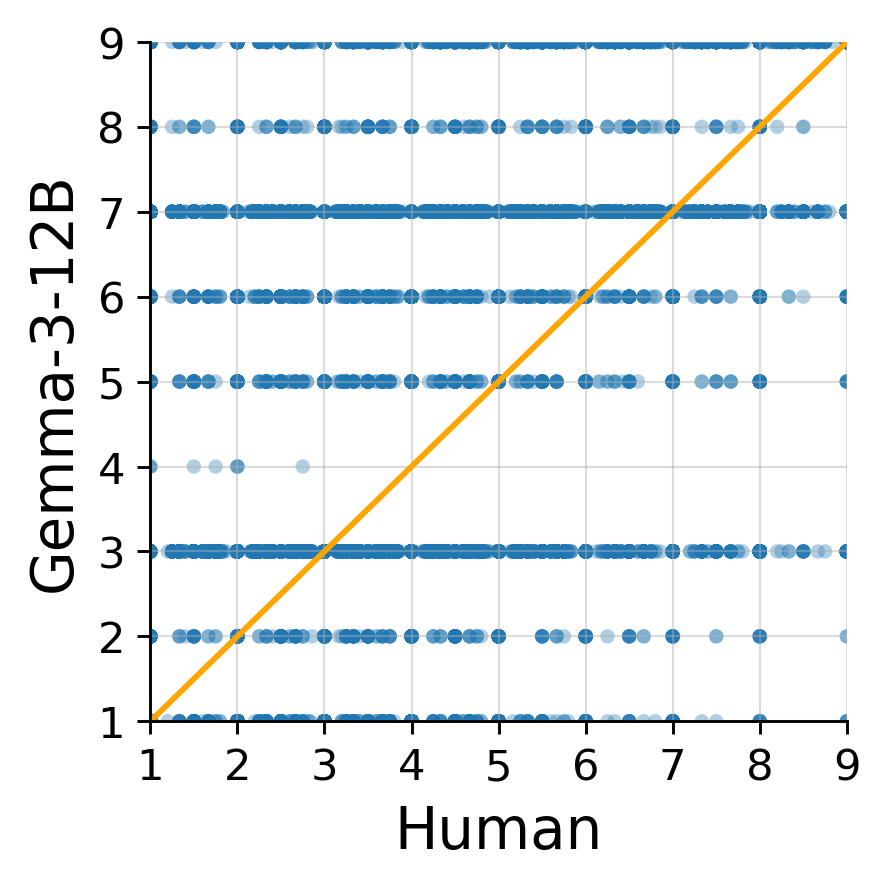}
\end{subfigure}\hfill
\begin{subfigure}[t]{0.19\textwidth}
  \centering
  \includegraphics[width=\linewidth]{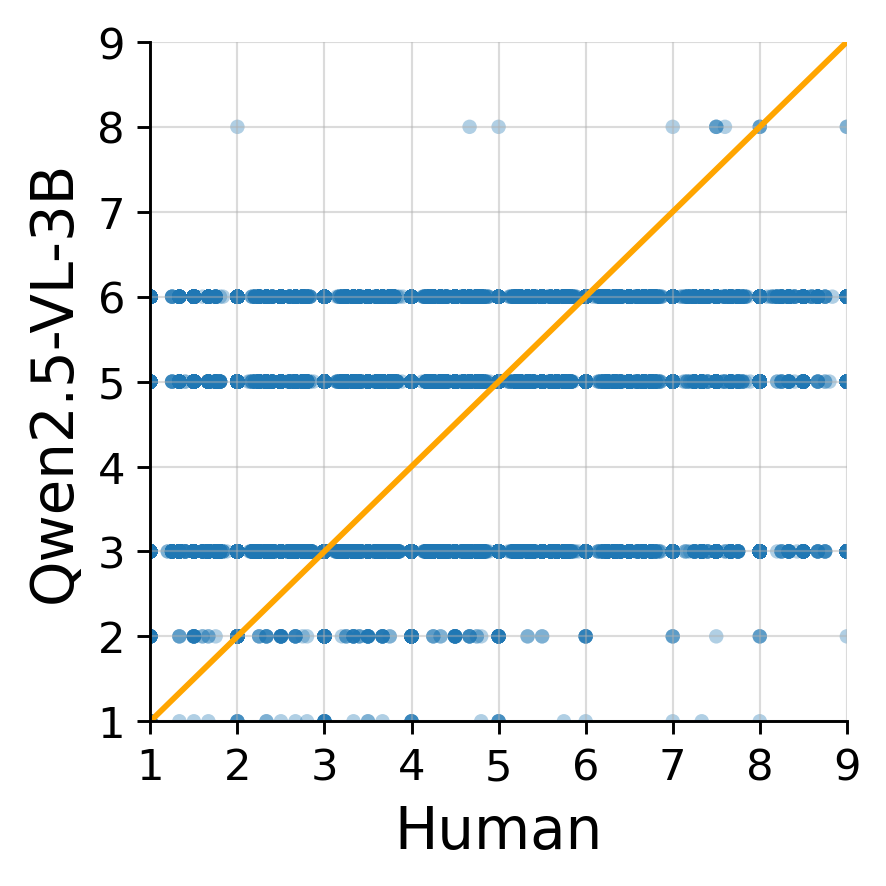}
\end{subfigure}\hfill
\begin{subfigure}[t]{0.19\textwidth}
  \centering
  \includegraphics[width=\linewidth]{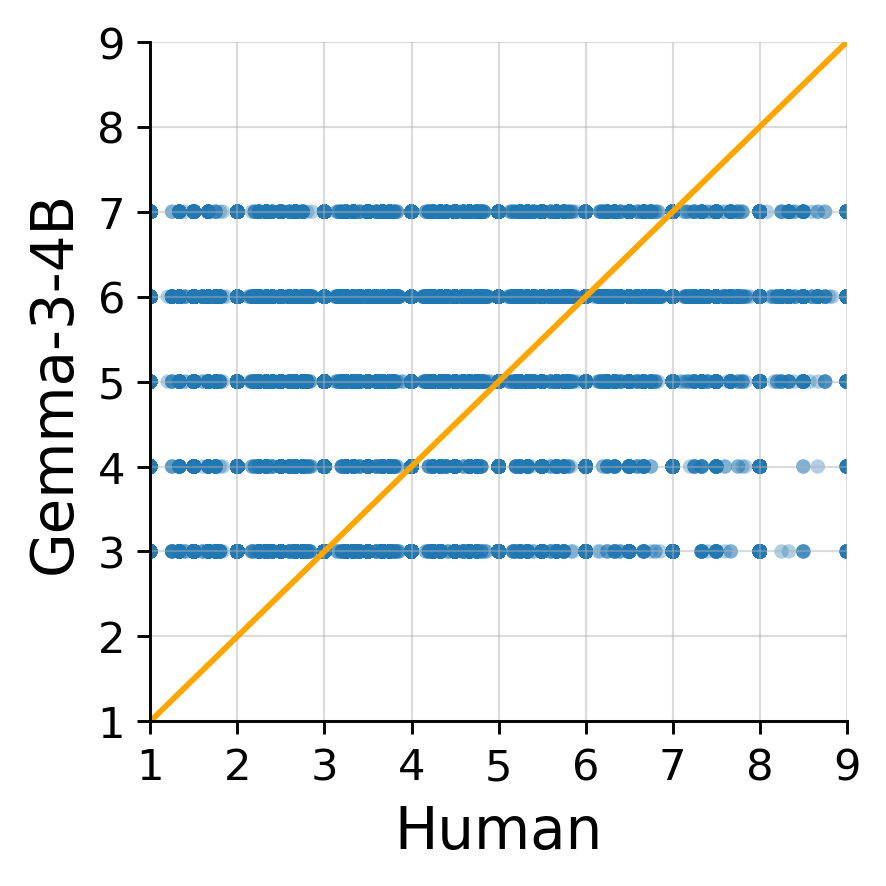}
\end{subfigure}

\caption{Individual-pair scatter plots (human vs.\ model) for VLMs.}
\label{fig:vlm_individual_scatter}
\end{figure}

\begin{table}[h!]
\centering
\caption{Procrustes correlations (Pearson $r$) for 10D GPT-4o MDS solutions rotated onto the human 8D space.}
\label{tab:procrustes_10d_gpt}
\setlength{\tabcolsep}{4pt}
\small
\begin{tabular}{lccccccccc}
\toprule
\multicolumn{1}{c}{Model (10D)} &
\multicolumn{1}{c}{\rot{D1: Lightness}} &
\multicolumn{1}{c}{\rot{D2: Grain}} &
\multicolumn{1}{c}{\rot{D3: Texture}} &
\multicolumn{1}{c}{\rot{D4: Shiny}} &
\multicolumn{1}{c}{\rot{D5: Organization}} &
\multicolumn{1}{c}{\rot{D6: Chromaticity}} &
\multicolumn{1}{c}{\rot{D7: Hue (R/G)}} &
\multicolumn{1}{c}{\rot{D8}} &
\multicolumn{1}{c}{\rot{Mean}} \\
\midrule
\rowcolor{vlmblue!10}
GPT-4o (baseline)  & 0.934 & 0.866 & 0.730 & 0.911 & 0.819 & 0.846 & 0.798 & 0.714 & 0.827 \\
\rowcolor{vlmblue!10}
GPT-4o (encourage middle) & 0.936 & 0.865 & 0.743 & 0.911 & 0.823 & 0.847 & 0.788 & 0.731 & 0.831 \\
\bottomrule
\end{tabular}
\end{table}


\begin{figure}[ht]
  \centering
  \includegraphics[width=\textwidth]{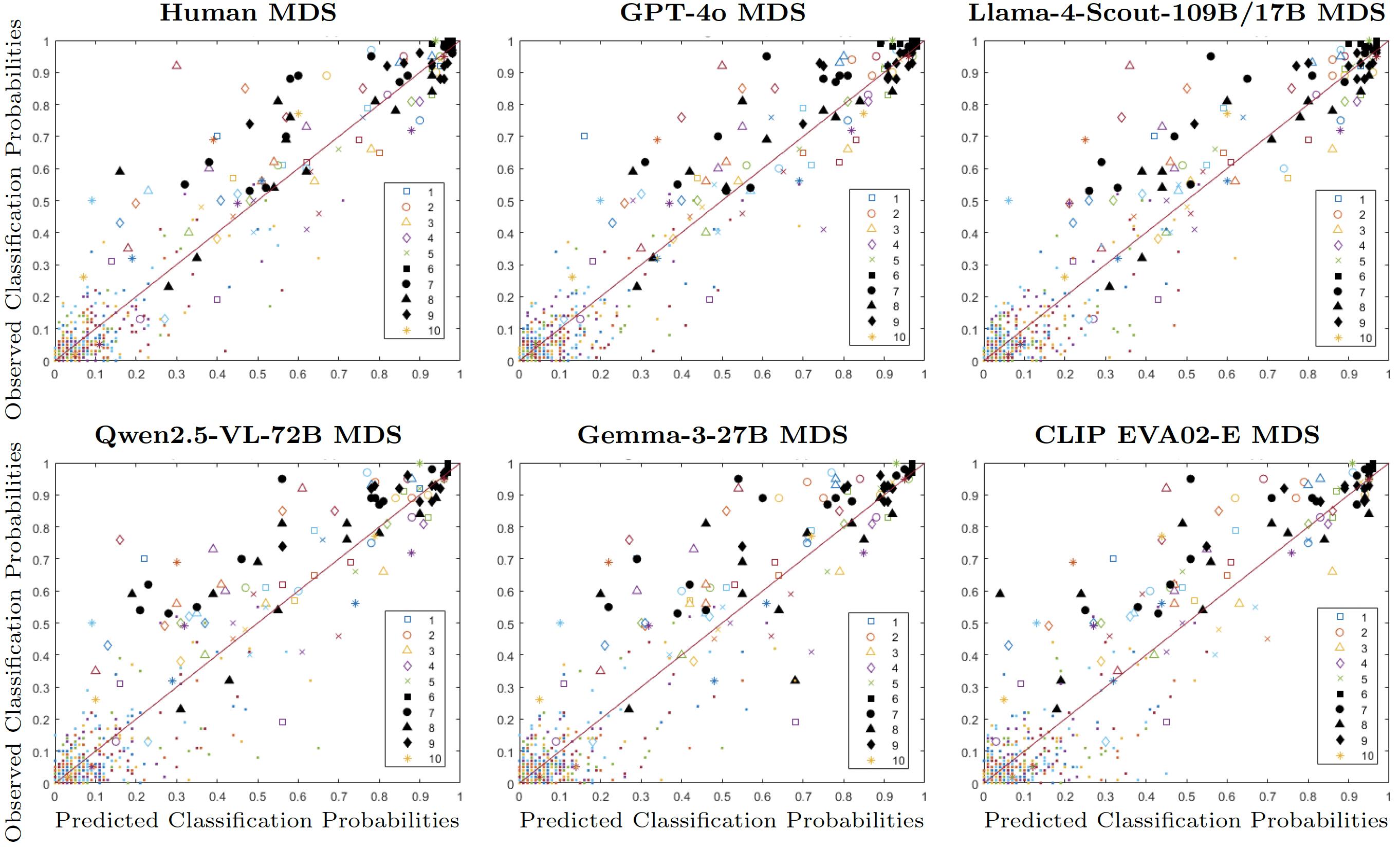}
  \caption{Predicted vs.\ observed classification probabilities for supplementary GCMs.}
  \label{fig:gcm_scatterplots_supl}
\end{figure}

\begin{figure}[ht]
\centering
\setlength{\tabcolsep}{3pt}
\begin{tabular}{cccc}
\includegraphics[width=0.24\textwidth]{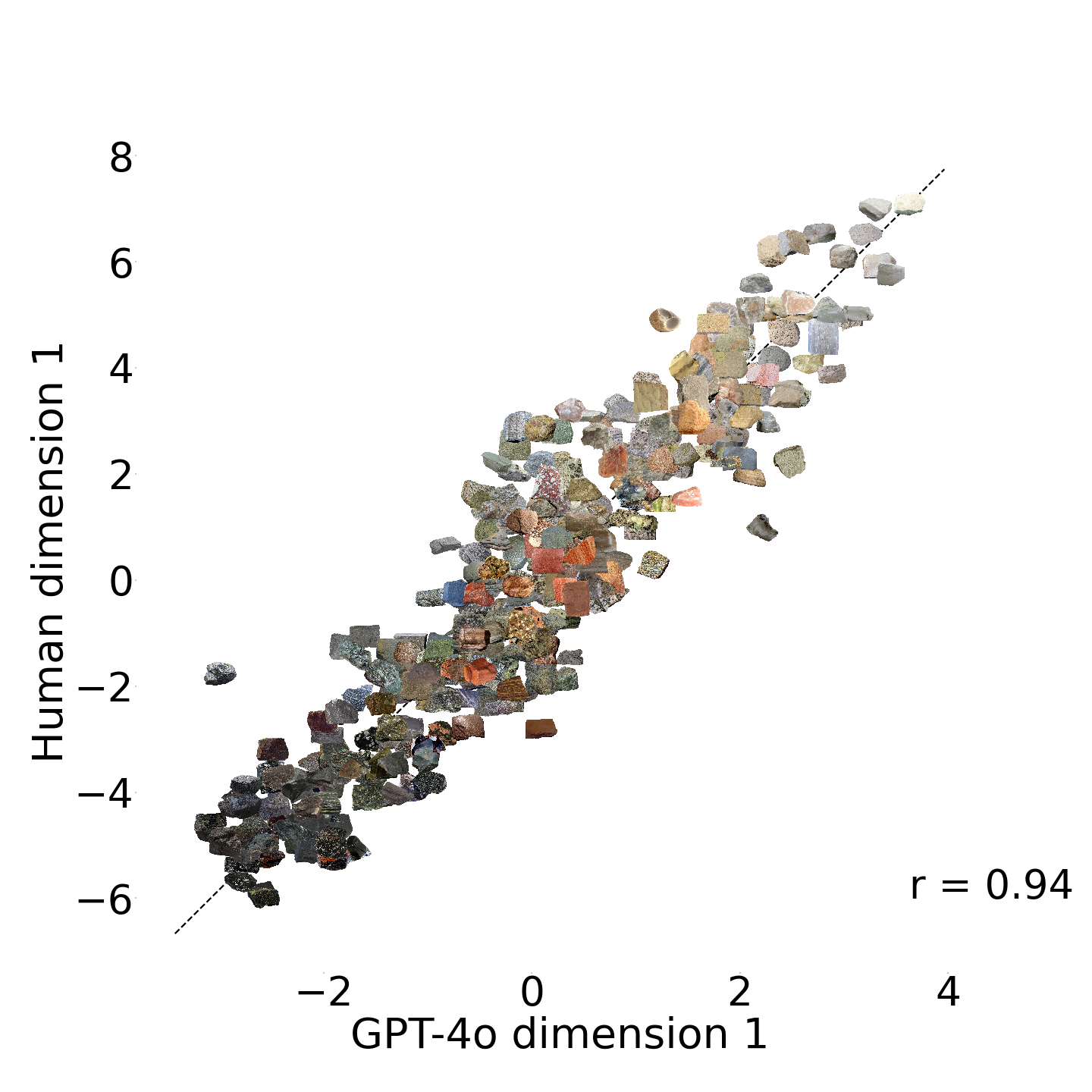} &
\includegraphics[width=0.24\textwidth]{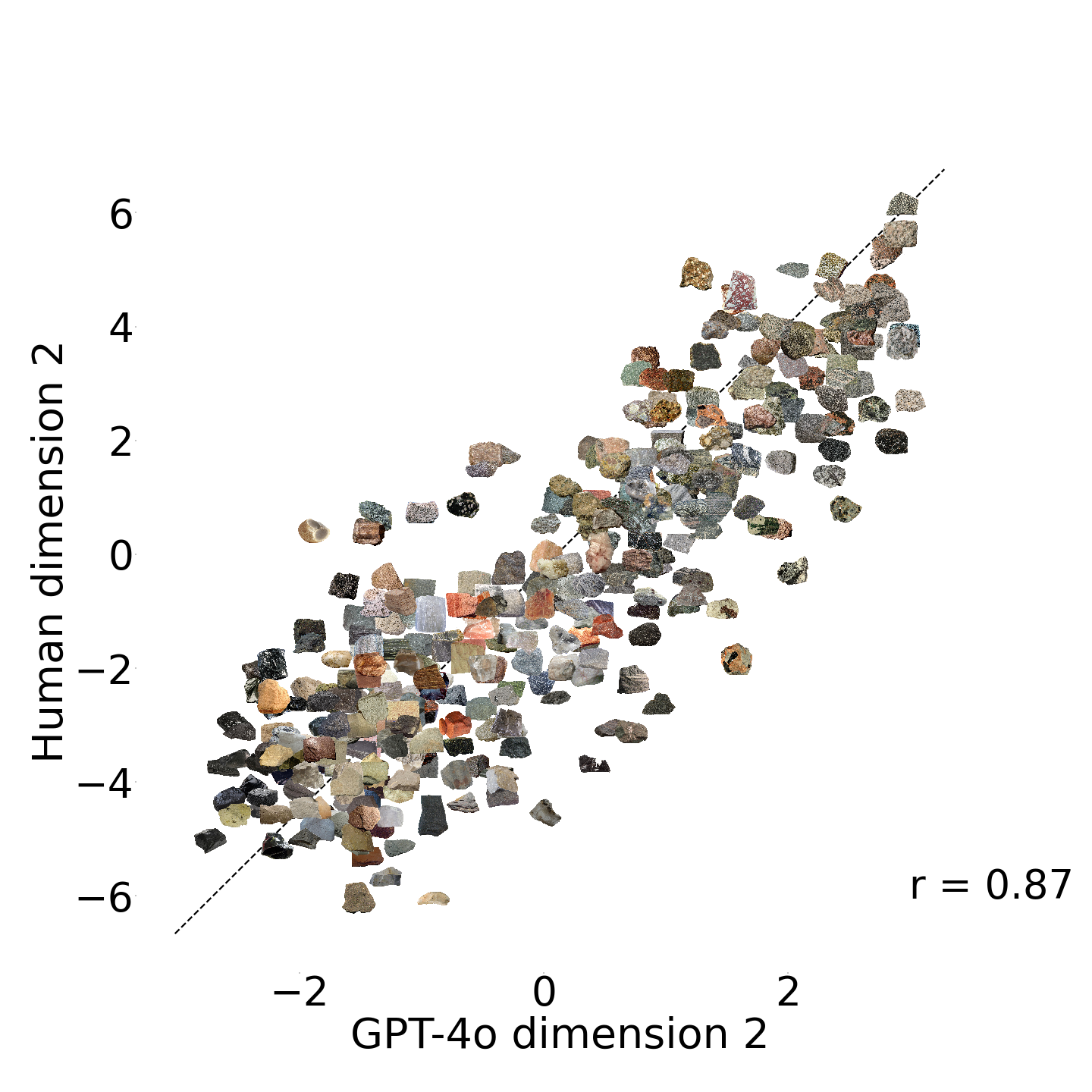} &
\includegraphics[width=0.24\textwidth]{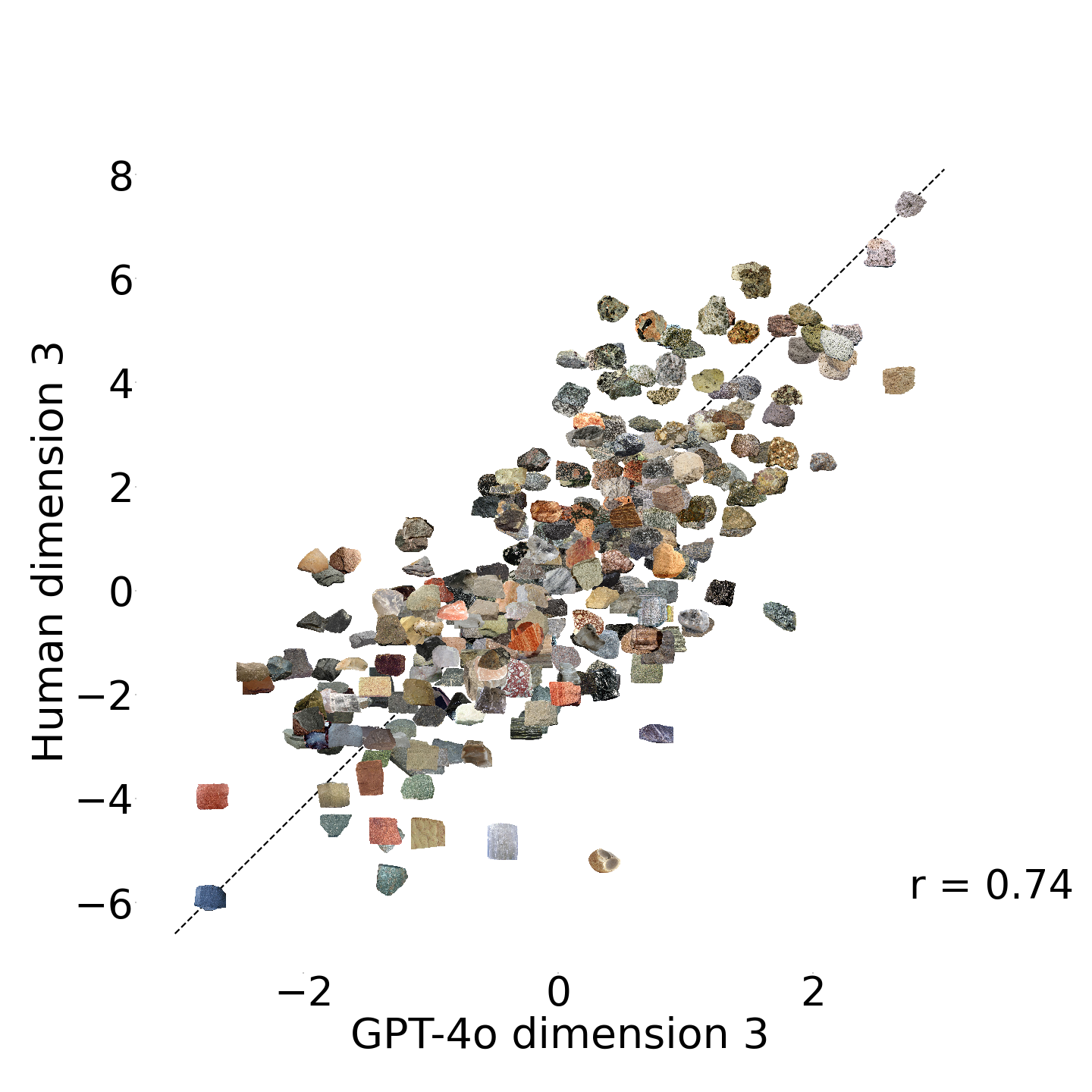} &
\includegraphics[width=0.24\textwidth]{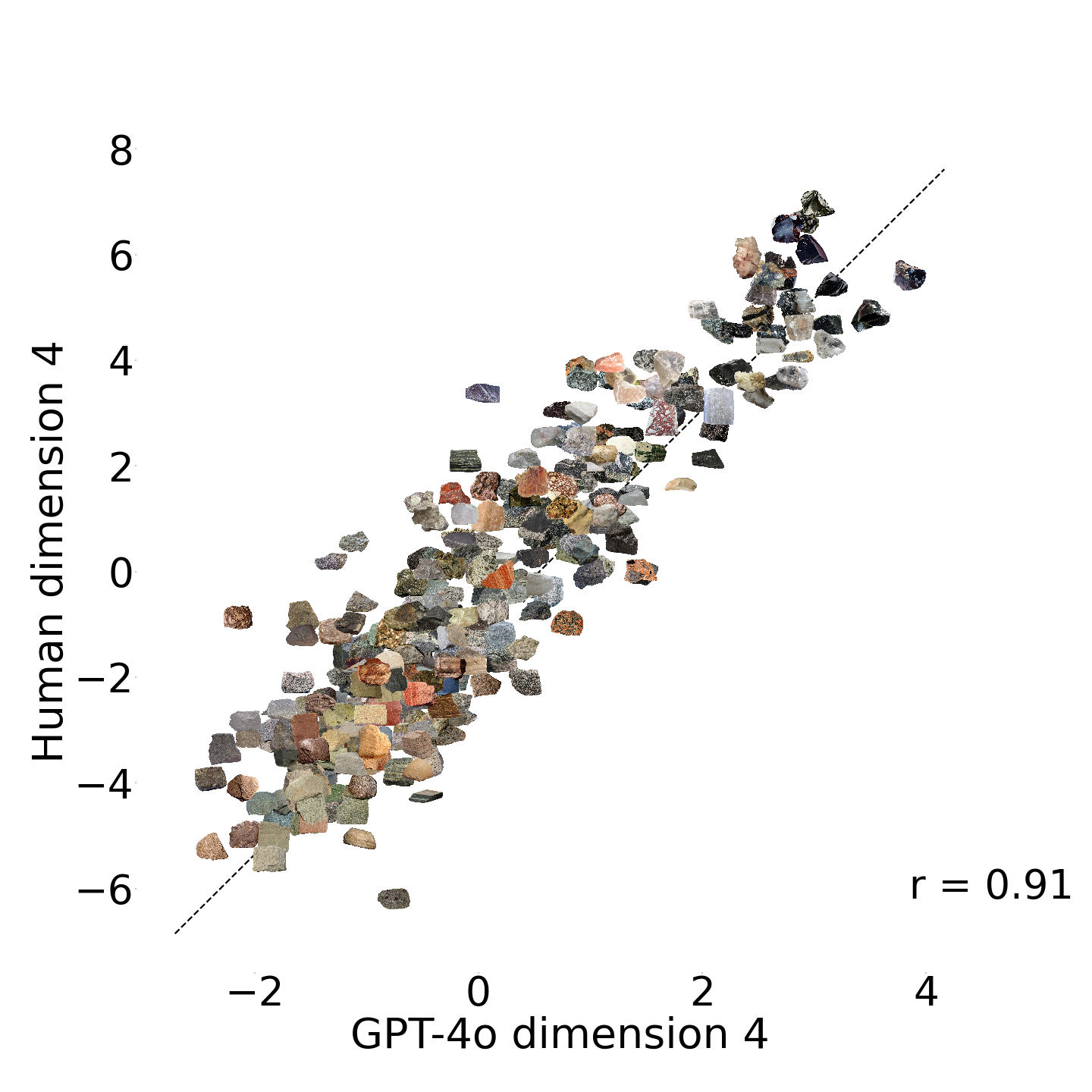} \\
\includegraphics[width=0.24\textwidth]{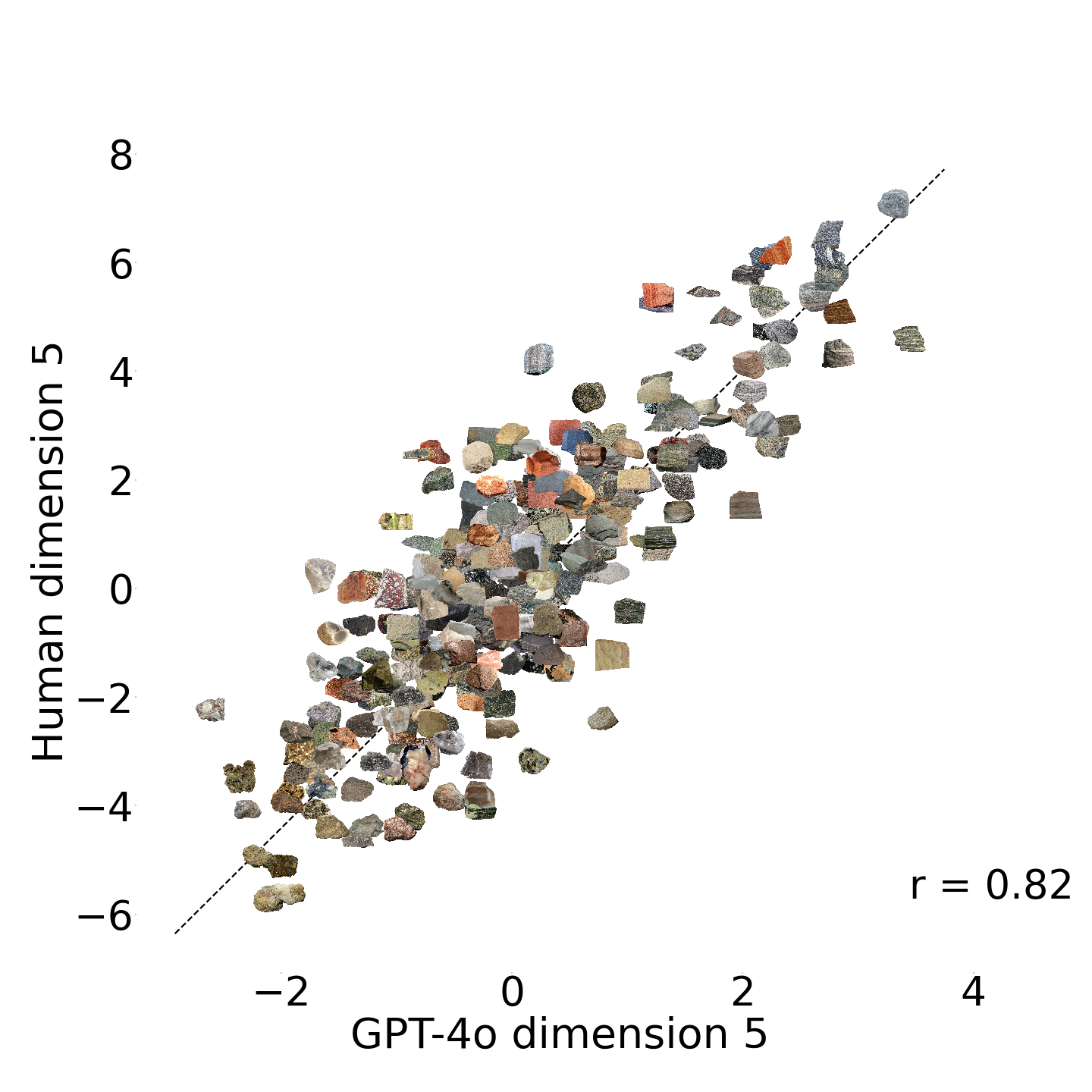} &
\includegraphics[width=0.24\textwidth]{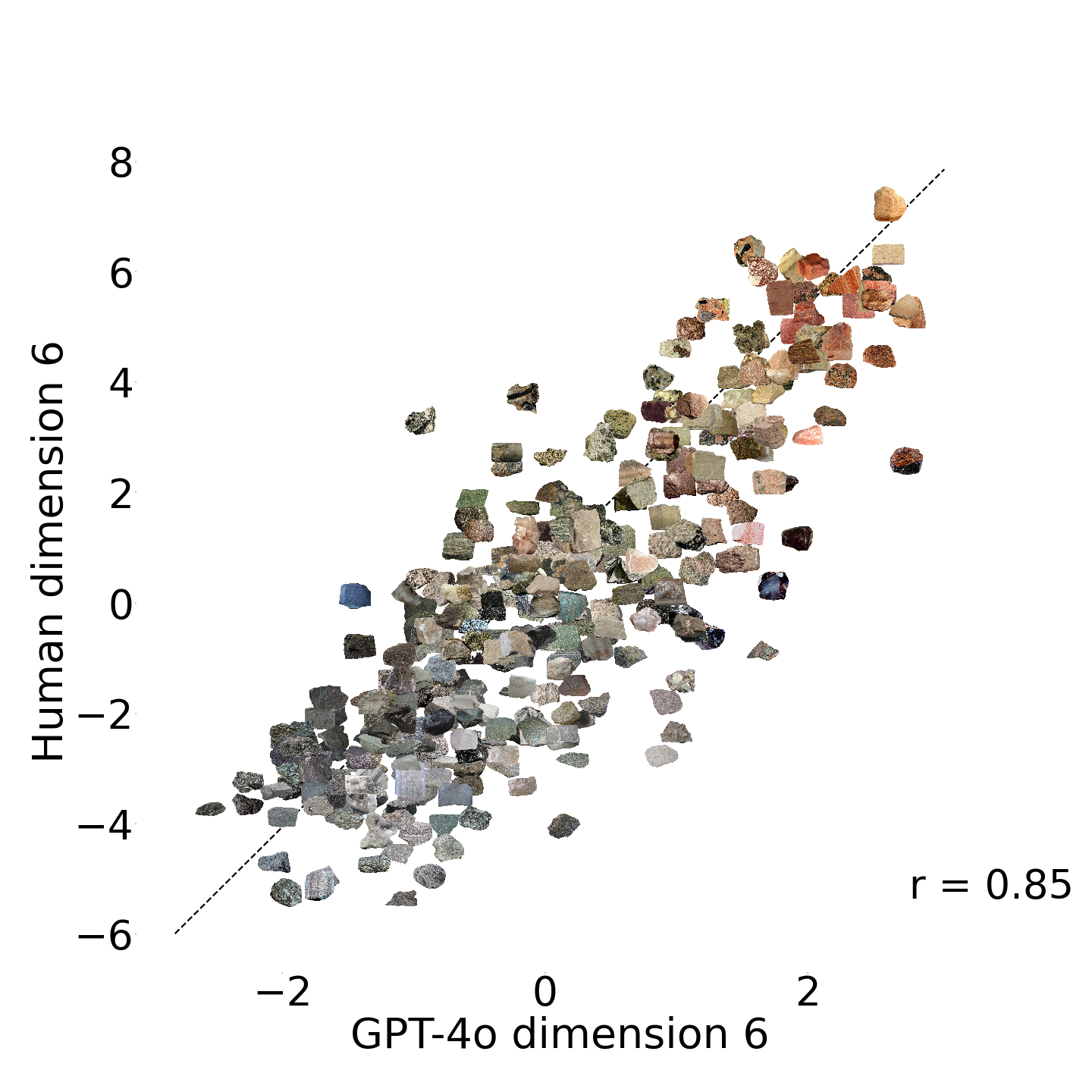} &
\includegraphics[width=0.24\textwidth]{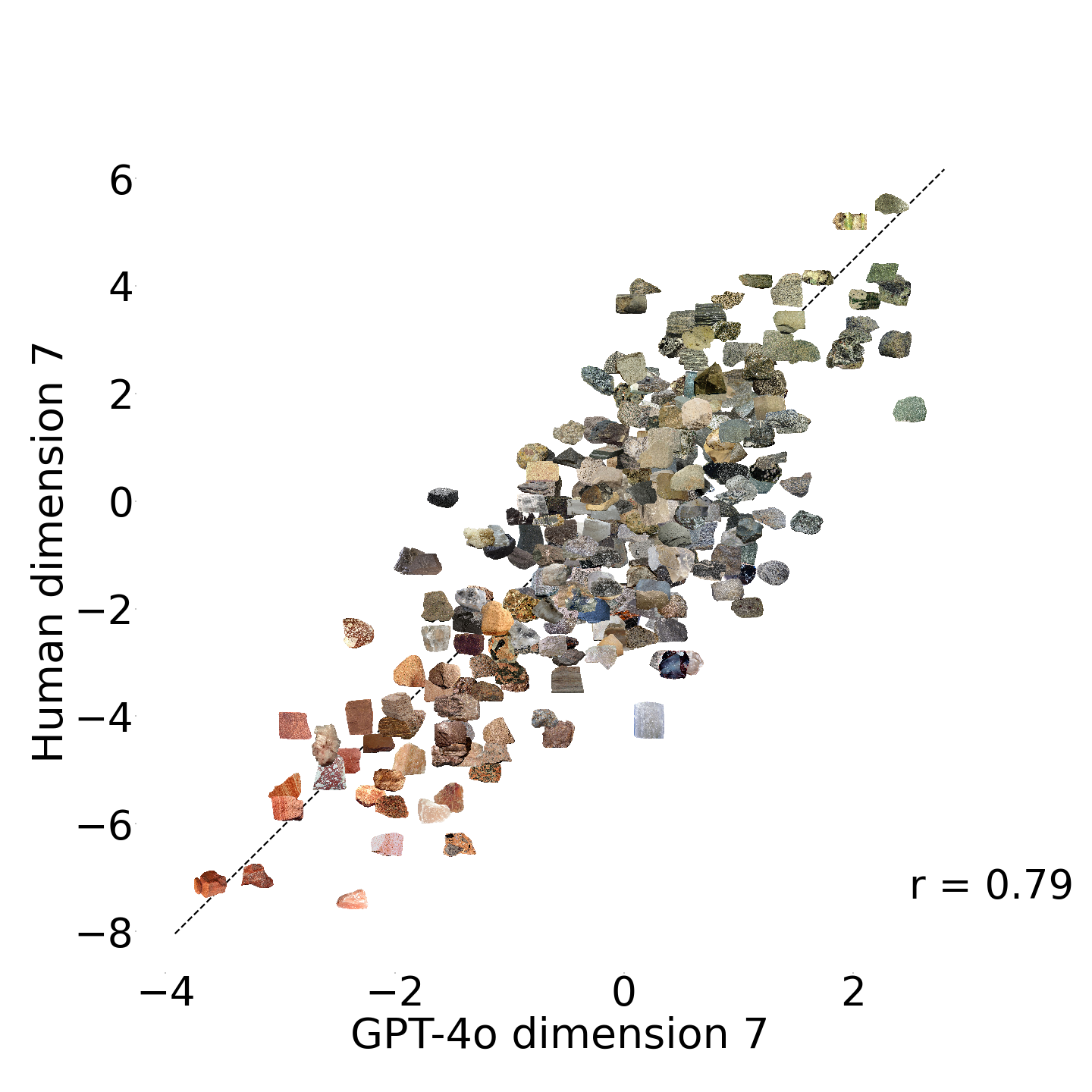} &
\includegraphics[width=0.24\textwidth]{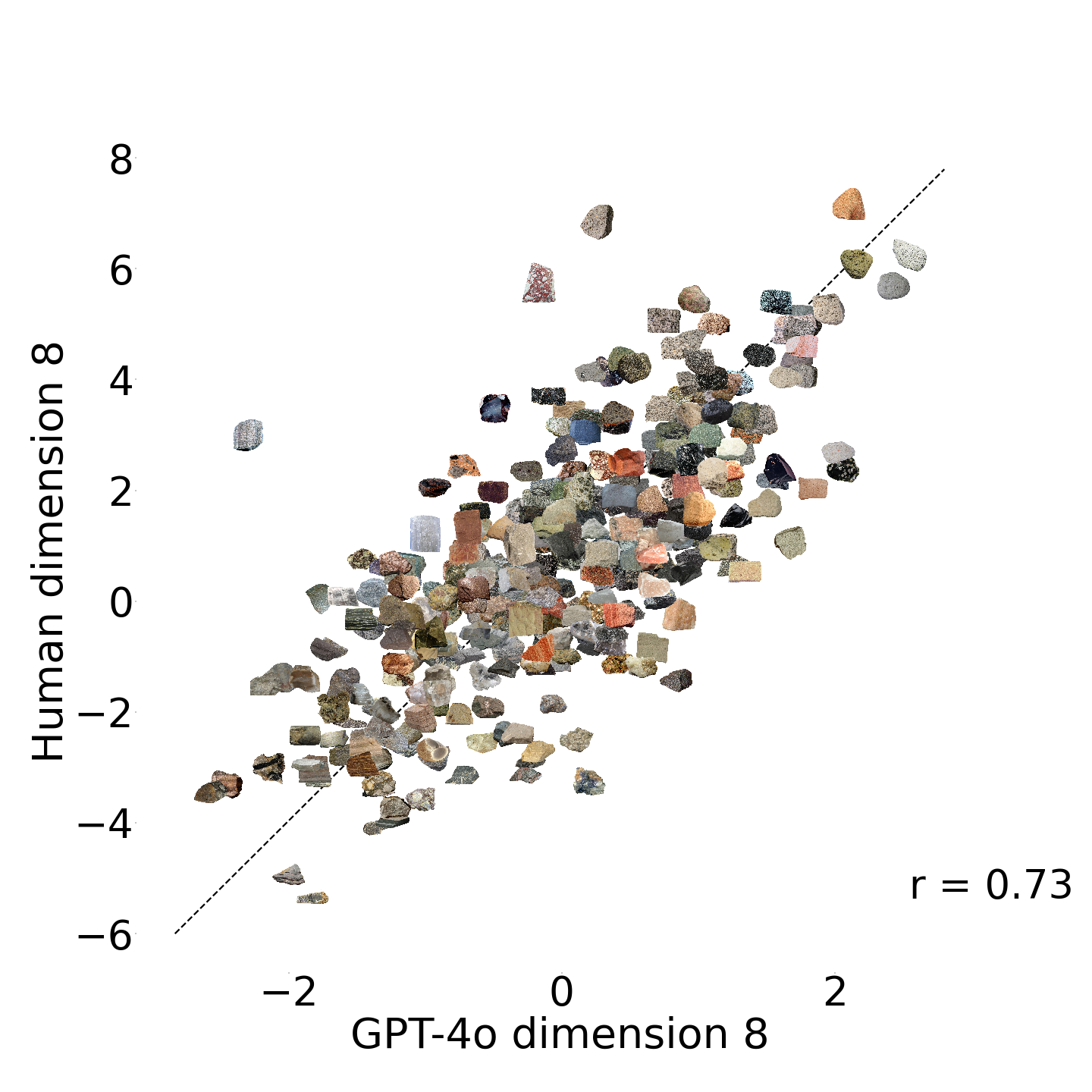} \\
\end{tabular}
\caption{Procrustes alignment between GPT-4o (encourage middle, 10D MDS) and the human 8D space, shown dimension by dimension. Each panel plots the aligned model coordinate (x-axis) against the corresponding human coordinate (y-axis) for all 360 rock images; the dashed line indicates equality and the in-panel $r$ is the Pearson correlation.}
\label{fig:gpt4o_procrustes_dims_10}
\end{figure}



\end{document}